\documentclass[conference]{IEEEtran}
\usepackage{geometry}                		
\usepackage[parfill]{parskip}    		        
\usepackage{graphicx}				
\usepackage{amssymb}				
\usepackage{amsmath}
\usepackage{textcomp}
\usepackage{breqn}
\usepackage{bbm}
\usepackage{tabularx}
\usepackage{latexsym}
\usepackage{subcaption}
\usepackage{gensymb}
\usepackage{amsmath}
\usepackage{fancyhdr}
\usepackage{eucal}
\usepackage{float}
\usepackage{parskip}
\usepackage{textcomp}

\usepackage{geometry}                		
\usepackage{tabularx}
\usepackage{stmaryrd}
\usepackage{wasysym} 
\usepackage{xurl}
\usepackage{multirow}
\usepackage[table]{xcolor} 
\newcommand{\code}[1]{\texttt{#1}}

\usepackage{microtype}
\usepackage{yfonts}
\usepackage{listings} 
\usepackage{multirow}
\usepackage{xcolor}
\usepackage{pdfpages}
\usepackage{graphicx} 
\usepackage[utf8]{inputenc}
\usepackage [english]{babel}
\usepackage [autostyle, english = american]{csquotes}
\usepackage[style=numeric,natbib=true]{biblatex} 
\graphicspath{{./Images/}} 
\usepackage{mathtools}
\usepackage{amsmath}
\usepackage{fancyhdr} 
\usepackage{titlesec}
\usepackage{upgreek} 
\usepackage{amsmath} 
\usepackage{amssymb}

\PassOptionsToPackage{hyphens, spaces, obeyspaces}{url} 
\usepackage[final=true,backref=true,colorlinks=true,linkcolor=blue, urlcolor=blue, citecolor=blue, breaklinks=true]{hyperref}
\hypersetup{
    pagebackref=true,
    colorlinks=true,
    citecolor=blue,
    linkcolor=blue,
    filecolor=magenta,      
    urlcolor=blue,
}
\hypersetup{breaklinks=true}
\usepackage{breakurl}

\def\xHyphenate#1#2\wholeString {\if#1$%
    \else\transform{#1}%
    \takeTheRest#2\ofTheString\fi}

\def\takeTheRest#1\ofTheString\fi
{\fi \xHyphenate#1\wholeString}

\def\transform#1{\url{#1}\hskip 0pt plus 1pt}

\definecolor{dkgreen}{rgb}{0,0.6,0}
\definecolor{gray}{rgb}{0.5,0.5,0.5}
\definecolor{mauve}{rgb}{0.58,0,0.82}
\definecolor{lightsilver}{rgb}{0.96,0.96,0.98} 
\definecolor{blond}{rgb}{1, 0.98, 0.8}
\lstdefinestyle{R} {
backgroundcolor=\color{blond},
language=Python,
basicstyle=\tiny,
} 
\titleformat{\chapter}[display]
{\normalfont\Huge\bfseries\raggedright}{\chaptertitlename\ 3}
{15pt}{\Huge}

\author{Justin London}
\begin{document}
\title{Biomedical Signal Processing: EEG and ECG Classification with Discrete Wavelet Transforms, Energy	Distribution, and Convolutional Neural Networks}
\author{\IEEEauthorblockN{Justin London}
\IEEEauthorblockA{\textit{Department of Electrical Engineering and Computer Science} \\
\textit{University of North Dakota}\\
Grand Forks, North Dakota USA \\
justin.london@und.edu}
}
\maketitle

\begin{abstract}
    Biomedical signal processing extract meaningful information from physiological signals like electrocardiograms (ECGs), electroencephalograms (EEGs), and electromyograms (EMGs) to diagnose, monitor, and treat medical conditions and diseases such as seizures, cardiomyopathy, and neuromuscular disorders, respectively.  Traditional manual physician analysis of electrical recordings is prone to human error as subtle anomolies may not be detected.  Recently, advanced deep learning has significantly improved the accuracy of biomedical signal analysis.  A multi-modal deep learning model is proposed that utilizes discrete wavelet transforms for signal pre-processing to reduce noise.  A multi-modal image fusion and multimodal feature fusion framework is utilized that converts numeric biomedical signals into 2D and 3D images for image processing using Gramian angular fields, recurrency plots, and Markov transition fields. In this paper, deep learning models are applied to ECG, EEG, and human activity signals  using actual medical datasets, brain, and heart recordings.  The results demonstrate that using a multi-modal approach using wavelet transforms improves the accuracy of disease and disorder classification.   
\end{abstract} 

\section{Introduction}

\ \ \ Electroencephalography (EEG) is the recording of electrical signals emanated from human brain, which can be collected non-invasively from the scalp by amplifying voltage differences between electrodes placed on the scalp or cerebral cortex and a brain-computer interface (BCI). A brain-computer interface (BCI) is a direct communication and control system that is established between the human brain and an electronic device.   The number of electrodes can vary from one to 256. The electrodes are placed at certain predefined positions according to the international 10/20 system or its variants. The weak electrical activity detected by the electrodes ranges from 5 to 100 $\mu$V, and the frequency range of interest is between 1 - 40 Hz \cite{Edlinger:2023}.  EEG converts \enquote{brain activity to time series data with amplitude on the y-axis, and this data can then be used to understand brain functions. \cite{Vaibhav:2023}}  Figure \ref{fig:nodes} illustrates the placement of the electrodes on the scalp.
\begin{figure}[H]
    \centering      
    \includegraphics[width=0.75\columnwidth]{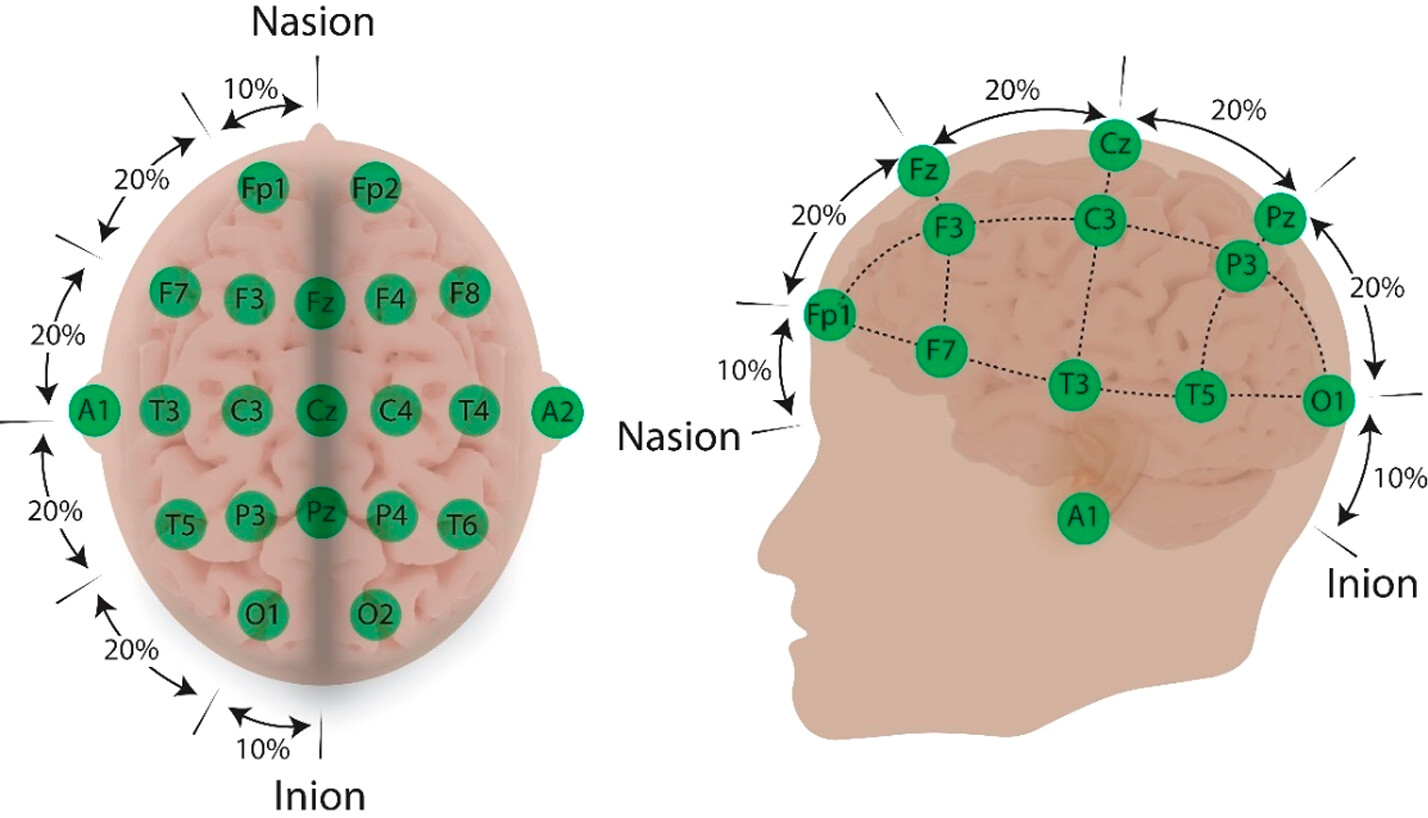} 
    \caption{Illustration of electrode arrangement placed on scalp. \cite{Kateb:2023}} 
    \label{fig:nodes}
\end{figure} 
Thus, each electrode has a number code that indicates where on the brain it is located.  The EEG measures voltage (difference in electrical potential) between each electrode and a reference electrode. Thus, whatever signal is present at the reference electrode is effectively subtracted from all the measurement electrodes.  The measurements generate for each electrode can serve as features of the EEG signal for deep learning.

\ \ \ Electrodes placed on the scalp in 3D space are \enquote{converted to a 2D space using azimuthal equidistant projection (AEP) to preserve the relative distance between electrodes present in the 3D space. The three scalar values, obtained per time frame, through the method of frequency binning, for each electrode, are interpreted as channels of an image, which are then projected onto a 2D electrode location. 

To generate smooth topographical 2D maps, Delaunay triangulation interpolation is utilized. This interpolation algorithm estimates the spectral measurement values in-between the electrodes over a 32 × 32 grid. Smooth reconstructed spatial maps are generated with different activation regions having fine transitions between them. \cite{Gopan:2022}}

Various electroencephalogram (EEG) signals have been used in BCI systems, such as P300 potentials \cite{Kshirsagar:2018, Gao:2015} , steady state visual evoked potentials (SSVEP) \cite{Kwak:2017}, and motor imagery (MI) \cite{Lu:2016, Ma:2017, Feng:2020}.  The recognition of MI-EEG is often challenging for several reasons. First, the \enquote{high-dimensional MI-EEG signal is too weak and its signal-to-noise ratio is low \cite{Müller:2008}. Secondly, the MI-EEG signal is a nonlinear and non-stationary signal, which means that its parameters, such as mean and variance change along with time \cite{Acharya:2005, Andrezjak:2001}. Further, MI signals are time-varying signals that depend on time variables \cite{Tang:2020}. In general, MI-EEG signals are highly complex and unstable signals, resulting in challenges for MI-EEG feature extraction and classification. \cite{Feng:2020}.}  

The most common method to classify EEG waveforms is by their frequency.  EEG waves are named based on their frequency range using Greek numerals. The most commonly studied waveforms include delta (0.5 to 4 Hz); theta (4 to 7 Hz); alpha (8 to 12 Hz); sigma (12 to 16Hz), beta (13 to 30Hz), and gamma (30-45 Hz).  Classification algorithms for EEG-based BCIs can be divided into four main categories: adaptive classifiers, matrix and tensor classifiers, transfer learning and deep learning, plus a few other miscellaneous classifiers \cite{Lotte:2018, Xie:2020}.   

\begin{figure}[H]
	\centering
	\includegraphics[width=0.75\columnwidth]{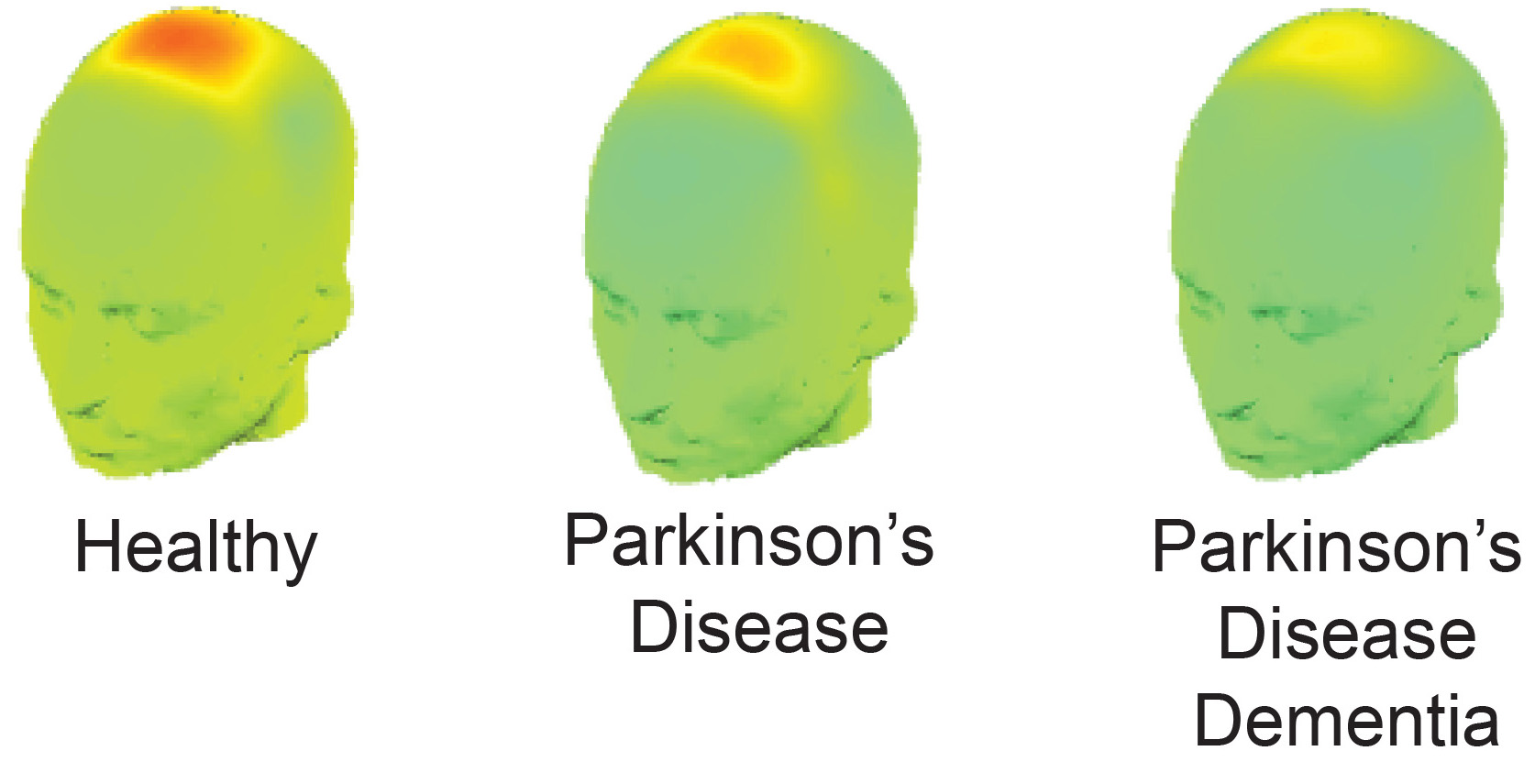} 
    \caption{Comparison of normal brain and Parkinson disease. \cite{Brown:2023}} 
    \label{fig:Parkinson}
\end{figure}

Electroencephalographic signals are used to diagnose and monitor seizure disorders as well as identify causes of other health problems.  Clinical applications of EEG signal processing include diagnosis and treat brain diseases as well as improve brain health (e.g., in epilepsy, neuropathic pain or mood disorders such as depression) and cognitive function (e.g., memory in dementia or Alzheimer's disease patients or executive function in children with ADHD).  EEG brain waves may predict cognitive impairment in Parkinson's disease.  For instance, Figure \ref{fig:Parkinson} illustrates the brain of a normal person and someone with Parkinson's.   
Some research has examined how using energy distribution of EEG signals as input into neural networks can accurately differentiate distinct motor movements of paralyzed individuals via thought controlled robotic devices and artificial limbs \cite{Contreras:2014, Avdakovic:1994} 

EEG signals can be used to diagnose sleep disorders and changes in behavior. They are sometimes used to evaluate brain activity after a severe head injury or before a heart transplant or liver transplant.  EEGs are often recommended for individuals with neurodevelopmental conditions like autism, intellectual developmental disorder, or global developmental delay \cite{Reyes:2022}.  EEGs are also used in brain computer-interface (BCI), neuroscience, and emotion detection \cite{Vaibhav:2023, Kumar:2022}.   

The International League Against Epilepsy (ILAE) defines a seizure as \enquote{a transient occurrence of signs and/or symptoms due to abnormal excessive or synchronous neuronal activity in the brain.}  Epileptic seizures occur at such high rates in autism and other neurodevelopmental disorders because disturbance to the brain that leads to developmental delays also predisposes the brain to develop seizures. 

\begin{figure}[H]
	\centering
	\includegraphics[width=0.75\columnwidth]{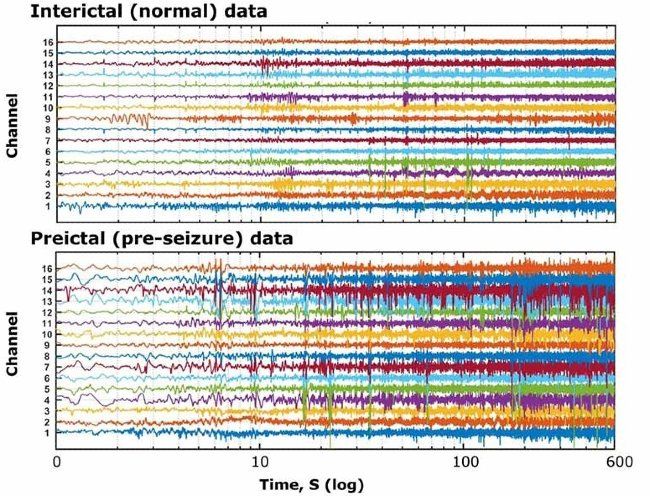} 
    \caption{Comparison of interictal (normal) brain and preictal (pre-seizure).  \cite{Jones:2023}}
    \label{fig:epile}
\end{figure}

Epileptic seizures \enquote{resulting from sudden excessive electrical discharges in a group of neurons may be accompanied by loss of awareness or consciousness and disturbances of movement, sensation, mood, or mental function \cite{Yuan:2011}.}  In some cases, severe seizures may also disrupt the brain in ways that lead to developmental delays. In those cases, the reduction or elimination of seizures through brain surgery and other treatments can be initially diagnosed through EEG signal processing \cite{Faust:2015}.   Figure \ref{fig:epile} illustrates EEG signals for a normal (interictal) person with a person who is preictal (pre-seizure).

In seizure diagnosis signal preprocessing is used to \enquote{remove artifacts occurred by blinking, wandering baseline (electrodes movement) and eyeball movement using the Discrete Wavelet Transformation (DWT). De-noising EEG signals from the AC power supply frequency with a suitable notch filter is another job of preprocessing. \cite{Arab:2010}} In experimental data, Arab, et. al. \cite{Arab:2010} found that the preprocessing enhanced speed and accuracy of the processing stage (wavelet transform and neural network). The EEGs signals are categorized to normal and petit mal and clonic epilepsy by an expert neurologist. The categorization is confirmed by Fast Fourier Transform (FFT) analysis and brain mapping.  

EEG can also be used to capture waves from visualization.  EEG signals can be filtered and processed in real-time to remove the ocular, muscle, and movement artifacts, as well as the line noise.  The EEG bandpower in different frequency-bands is extracted.
This information is used to calculate various features from emotions evoked from visualization of different images/videos such as valence, arousal, engagement, attention, and fatigue \cite{Kwak:2017}.  

Just as an EEG can be used to analyze electrical signals generated from brain waves, ECGs can be used to analyze electrical signals generated from heart beats. According to published medical reports, cardiovascular disease remains the leading cause of mortality worldwide.   Using 12 lead electrodes placed on the patient's chest, an electrocardiogram (ECG), also known as an EKG, records the heart's electrical activity through repeated cardiac cycles.  

One of the main causes of cardiovascular diseases is cardiac arrhythmia, in which the heartbeat deviates from typical beating patterns. There are different types of heart arrhythmia including premature ventricular contraction (PVC), atrial premature contraction (APC), right bundle block (RBBB), and left bundle branch block (LBBB) \cite{Subasi:2019}.   

\begin{figure}[H]
   \centering	\includegraphics[width=0.75\columnwidth]{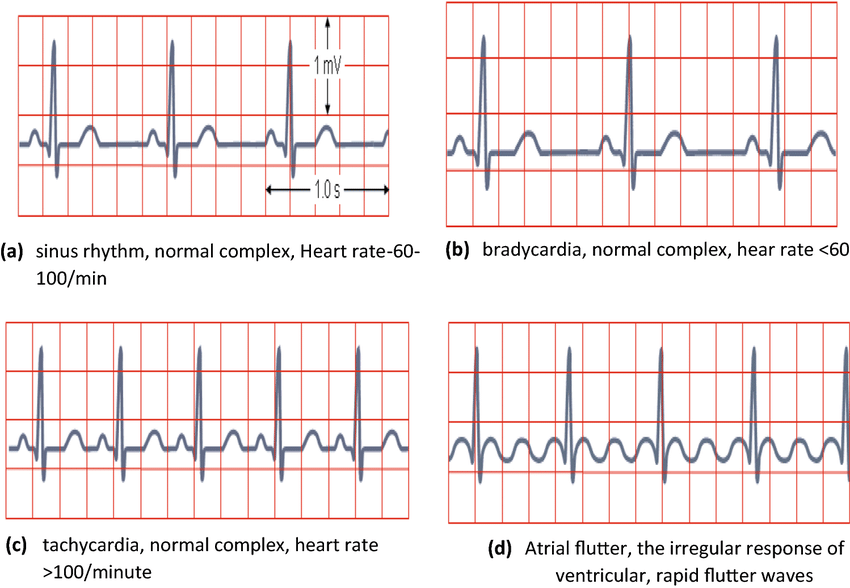} 
    \caption{Illustration of various normal and irregular ECG signals.  \cite{Thir:2021}}
    \label{fig:ecg}
\end{figure}

ECGs can be used to investigate symptoms of a possible heart problem, such as chest pain, palpitations (suddenly noticeable heartbeats), dizziness and shortness of breath. Thus, an ECG can help detect: (1) arrhythmia (where the heart beats too slowly, too quickly, or irregularly); (2) coronary heart disease (where the heart's blood supply is blocked or interrupted by a build-up of fatty substances(; (3)  heart attacks (where the supply of blood to the heart is suddenly blocked); and (4) cardiomyopathy (where the heart walls become thickened or enlarged.)  Figure \ref{fig:ecg} illustrates various ECG signals.

Feature extraction and feature selection are important steps in the process of EEG and ECG signal classification. The quality of the features affects the accuracy of the classifiers
(Begg et al., 2008). One of	the	major advantages to	using the DWT over other types	of	transforms	such as	the	Fourier	 transform	is	that the DWT is	able to	achieve	 both temporal	and spatial resolution while others  capture just frequency resolutions.  Thus, the	DWT	is	able to	capture	both the frequency and location of the signal.  Deep learning neural networks have become  powerful tools to extract features and to classify biomedical signals \cite{Anwar:2020}, \cite{Matlab:2023}, \cite{Craik:2019}, \cite{Smigiel:2021}.  Figure \ref{fig:process} illustrates the process of classification of EEG signals:
\begin{figure}[H]
    \centering
\includegraphics[width=0.75\columnwidth]{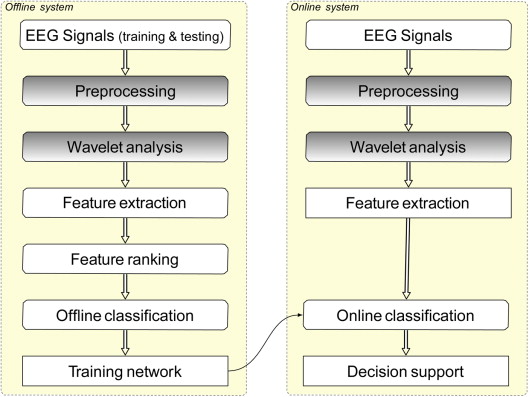} 
    \caption{Biomedical signal processing diagram.}
    \label{fig:process}
\end{figure} 

\ \ \ While multiple representations of EEG can be used such as 1D (signals or feature vector), 2D feature maps (images), and 3D (video).  In this project, we use 1D signals which are the most commonly used.  However, we show that we can use 1D signal to generate 2D scalogram feature map images from the wavelet transform which can in turn be input into convolutional neural networks and thereby improving classification accuracy of EEG signals types.  

In this paper, we analyze three types of biomedical signals, EEG and ECG using wavelet transforms to extract meaningful features for classification and detection of epilepsy and cardiac rhythmic abnormalities, respectively.  We also analyze human activity recognition (HAR) signals which measure human activity using accelerometer and gyroscope sensors.

Wavelets serve an important purpose as not only due they aid in denoising the signal via cascading (upsampling and downsampling) filtered  banks, but they enable the transformation of time series data into images that can be used to extract meaningful signal features.  We also review the use of multi-modal algorithm methods used by researchers such as Gramian Angular Fields (GAFs) that, like wavelet transforms, enable the conversion of 1D signals to 2D images and can better capture non-linear and non-stationary dynamics to improve biomedical signal classification and therefore detection and diagnoses of brain and heart diseases.   

\section{Methods}

\ \ \ In this project, EEG, ECGs and HAR signals will be examined using the discrete/continuouis wavelet transform (DWT)/(CWT), energy distributions, and neural networks to classify different EEG signals.  Since EEG signals are so important in detecting brain diseases, it is critical to properly preprocess them for accurate assessment and detection.  Likewise, it is critical to properly preproess ECG signals to detect heart disease and irregularities. 

This project will analyze use of the discrete wavelet transform process and convolutional neural networks to process and classify EEG and ECG biomedical signals.  MI-EEG data will be used to classify and analyze epilepsy and seizure disorders.  Wavelet transforms will be used to denoise and compress biomedical signals in the preprocessing step to improve the classification accuracy of the CNN.  The Digital Signal Processing and Wavelet Processing Matlab Toolboxes will be used in addition to the Deep Learning Toolbox and TensorFlow/PyTorch.

Various EEG and ECG datasets will be used.  The PhysioNet database will be used for EEG signals from epileptic patients\footnote{\url{https://physionet.org/content/siena-scalp-eeg/1.0.0/}. See also \url{https://physionet.org/content/?topic=eeg}.}  The MIT-BIH Arryhtmia Database for ECG data\footnote{\url{https://www.physionet.org/content/mitdb/1.0.0/}} will be used as well as the Mendeley Dataset\footnote{\url{https://data.mendeley.com/datasets/7r4z3p3g4m/1}} for EEG and ECG.  The later dataset includes the data with noise and without noise.

Algorithms discussed in Feng et. al. \cite{Feng:2020} that combine continuous wavelet transform (CWT) and a simplified convolutional neural network (SCNN) to improve the recognition rate of MI-EEG signals will be used. Algorithms in \cite{Subasi:2019}, \cite{Xin:2022}, and \cite{Cui:2016} will also be used.  

\section{Wavelet Transforms}

\ \ \ Wavelets are short-duration waveforms or mathematical functions that are localized in time and frequency, but can be scaled and translated to analyze data in different frequency ranges and locations.  Wavelets have two basic properties: scale and location.  Scale (or dilation) define how \enquote{stretched} or \enquote{compressed} a wavelet is and is related to wave frequency. 

Location define where the wavelet is positioned in time (or space).  Wavelet transforms are powerful filtering processes for multiresolution analysis by decomposing a signal over different scale levels. A wavelet must have (1) finite energy and (2) zero mean.  Finite energy means that it is localized in time and frequency and is integrable and the inner product between the wavelet and and the signal always exists.

Wavelets decompose signals into time-varying frequency (scale) components. Since signal features are often localized in time and frequency, analysis and estimation are easier when working with sparser (reduced) representations.  For instance, the wavelet can decompose an EEG signal into five subbands: delta (0–4 Hz), theta (4–8 Hz), alpha (8–15 Hz), beta (15–30 Hz), and gamma (30–60 Hz).  

Scales can be converted into pseudo-frequencies via the equation $f_{a} = f_{c}/a$ where $f_{a}$ is the pseudo-frequency, $f_{c}$ is the central frequency of the mother wavelet, and $a$ is the scaling factor \cite{Taspinar:2018}.  High scale-factors (longer wavelets) correspond with a smaller frequency.  Thus, by scaling a wavelet in the time-domain, one can analyze smaller frequencies achieving higher resolution.  Conversely, by using smaller scales, one has more detail in the time-domain.

The Fourier transform has a high resolution in the frequency domain and zero-resolution in the time-domain.  The STFT has medium sized resolution in both the frequency and time domain.  For small frequency values, the wavelet transform has a high resolution in the frequency domain and low resolution in the time-domain.  For large frequency values, the wavelet transform has low resolution in the frequency domain and high resolution in the time domain \cite{Taspinar:2018}. 

\subsection{Discrete Wavelet Transform}

The discrete wavelet transform (DWT) is used in image compression, feature extraction, and denoising and for non-stationary signals \cite{Zaid:2020}.   The DWT can be used to reduce image noise by removing high-frequency components by thresholding the wavelet coefficients.  Since wavelets localize features in the data to different scales, one can preserve important signal or image features while removing noise. 

The basic idea behind wavelet denoising, or wavelet thresholding, is that the wavelet transform leads to a sparse representation for many real-world signals and images. In other words, the wavelet transform concentrates signal and image features in a few large-magnitude wavelet coefficients. Wavelet coefficients which are small in value are typically noise and one can \enquote{shrink} those coefficients or remove them without affecting the signal or image quality. After thresholding the coefficients, one can reconstruct the data using the inverse wavelet transform.
\begin{equation}
W_{\phi}(j,k) = \frac{1}{\sqrt{M}}\sum_{x} f(x) \psi_{j,k}(x)
\label{eq:eq1}
\end{equation}
The inverse DWT is defined as
\begin{multline}
    f(x) = \frac{1}{\sqrt{M}}\sum_{k} W_{\phi}(j_{0},k)\phi_{j_{0},k}(x) \\ + \frac{1} {\sqrt{M}} \sum_{j=j_{0}}^{M} \sum_{k} W_{\psi}(j,k)\psi_{j,k}(x)
    \label{eq:eq2}
\end{multline}
where $f(x)$, $\phi_{j_{0},k}(x)$, and $\psi_{j,k}(x)$ are functions of the discrete variable $x=0,1,2,\dots,M-1$.  Normally, $j_{0} = 0$ and $M = 2^{j}$ where the summations in equations \ref{eq:eq1} and \ref{eq:eq2} are performed over $x = 0,1,\dots,M-1$, $j=0,1,\dots,J$, and $k = 0,1,\dots,2^{j}-1$.  The coefficients in equations \ref{eq:eq1} and \ref{eq:eq2} are known as \textit{approximation} and \textit{detail} coefficients (sub-signals), respectively.  $\phi_{j_{0},k}(x)$ is a member of the set of expansion funcitons derived from a scaling function $\phi(x)$ by translation and scaling using:
\begin{equation}
    \psi_{j,k}(x) = 2^{j/2}\psi(2^{j}x - k)
\end{equation}
    A signal passes through a high-pass filter $\phi(j_{0},k)$ and a low pass filter $\psi(j,k)$.  
    It is then decomposed into high frequency waveforms (details) via the high-pass filter and low frequency waveform components (approximation) via the .  

    In general, a common function for wavelets is:
\begin{equation}
    \psi_{s,t}(t) = \lvert s \rvert^{1/2} \psi \bigg [ \bigg (\frac{t - \tau}{s} \bigg ) \bigg ]
\end{equation}
where $s$ and $\tau$, $s \neq  0$, denotes the scale and translation parameters, and $t$ denotes time.   The continuous time wavelets can be expressed as:
\begin{equation}
    X_{WT}(\tau,s) = \frac{1}{\sqrt{\lvert s \rvert}} \int x(t) \psi \bigg (\frac{t-\tau}{s} \bigg )dt
\end{equation}
    There are two keys for using wavelets as general feature detectors:
    First, the wavelet transform separates signal components into different frequency bands enabling a sparser representation of the signal.  Second, one can often use a wavelet which resembles the feature you are trying to detect.  For instance, the 'sym4' wavelet resembles the QRS complex, which makes it a good choice for QRS detection \cite{rwave:2023}.  The success of the wavelet transform (WT) depends on the optimal configuration of its control parameters which are often experimentally set. However, the optimality of the combination of these parameters can be measured in advance by using the mean squared error (MSE) function.  Several powerful metaheuristic algorithms have been proposed to find the optimal WT parameters for EEG signal denoising which are harmony search (HS), $\beta$-hill climbing ($\beta$-hc), particle swarm optimization (PSO), genetic algorithm (GA), and flower pollination algorithm (FPA) \cite{Zaid:2020}.   

    In general, DWT decomposes a signal by using a filter bank, a set of cascading low pass and high pass filters,  to produce the approximation and details coefficients, respectively. The main objective of using DWT is to decompose the input signal via different coefficient levels to correct the high frequency of the input signals [48]. In other word, DWT decomposes the EEG signal into several frequency bands because it assumed that the artifacts will have large amplitudes in the respective frequency bands.  The denoising process involves three phases:
\begin{itemize}
    \item Decomposition phase:  Assuming the original EEG or ECG signal with $n$ samples x(t) = [x(1),x(2),,\dots,x(n)] is divided into three levels, and each level will be composed into two parts, approximation coefficients ($c_{A}$) and detail coefficients ($c_{D}$).  $c_{D}$ will be processed using a high-pass filter while $c_{A}$ will continue to be composed for the next level.
    \begin{align}
        c_{A_{i}}(t) &= \sum_{k=-\infty}^{\infty} c_{A_{i-1}}(k)\phi_{i}(t - k ) \\
        c_{D_{i}}(t) &= \sum_{k=-\infty}^{\infty} c_{D_{i-1}}(k)\Psi_{i}(t - k) 
    \end{align}
    where $c_{A_{i}}(t)$ and $c_{D_{i}}(t)$ denotes the approximation and detail coefficients of level $i$. $\Psi$ and $\phi$ are scaling and shifting functions, respectively.
    \item Apply thresholding:  A threshold value is defined for each level according to the noise level of the coefficient.
    \item Reconstruction phase:  The EEG or ECG denoised signal $x(t)$ is reconstructed using the inverse DWT (\code{idwt}) given by:
    \begin{multline}
        x(t)_{rec} = \sum_{k = -\infty}^{\infty} c_{A_{L}}\phi'_{i}(t-k) \\+ \sum_{i=1}^{L}\sum_{k = -\infty}^{\infty} c_{D_{i+1}(k)}\Phi'_{i}(t) 
    \end{multline}
    where $x(t)_{rec}$ denotes the reconstructed signal and $i$ denotes the decomposition level, $i=1,..,L$.  In this case $L = 3$. $\phi'_{i}$ and $\Phi'_{i}(t)$ denote the derivative of the scaling and translation functions. The EEG signal denoising process is given in Figure \ref{fig:denoise}:
    \begin{figure}[H]
    \centering  \includegraphics[width=0.75\columnwidth]{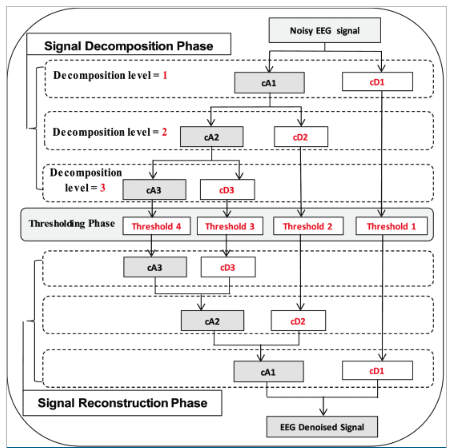} 
    \caption{EEG denoising process. \cite{Zaid:2020, Alyasseri:2018}}
    \label{fig:denoise}
\end{figure} 
\end{itemize}
    Wavelet transforms have five (denoising) parameters with each parameter: (1) the mother wavelet function $\Phi$ to be used (e.g. Symlet, Coiflet, Daubechies, and Biorthogonal), (2) thresholding function $\beta$ (soft or hard); (3) decomposition level $L$ (range of 5); (4) thresholding selection rule $\lambda$:
    \begin{itemize}
        \item Rigrsure : threshold is selected using the principle of Stein's unbiased risk estimate (SURE) 
        \item Sqtwolog : threshold is selected according to $\sqrt{2\text{log}M}$ 
        \item Heursure: threhsold is selected according to mixture of Rigrsure and Sqtwolog 
        \item Minimax: threshold is selected equal to max(MSE). 
    \end{itemize}
     and (5) re-scaling approach $\rho$ (one (no scaling), sln (single level), mln (multiple level)).  The thresholding mechanisms must be applied because the selection will affect global denoising performance \cite{Alyasseri:2018}.  The parameters can be selected through meta-heuristic algorithms such as a genetic algorithm, particle swarm optimization, harmony search algorithm, flower pollination algorithm, or $\beta$-hill climbing algorithm.


\section{Short Time Fourier Transform}

\ \ \ The Short Time Fourier Transform (STFT) overcomes the temporal limitations of the Fourier Transform. STFT performs a Fourier transform on on a small window of data at at time thereby mapping the signal into a 2D function of time and frequency.  The STFT can be represented as:
\begin{multline}
    X(k,m) = \sum_{n=0}^{N-1}x(n+m)w(n)W^{k}_{N}, \\  k,m = 0,1,\dots,N-1
\end{multline}
    The squared magnitude of the STFT yields the spectrogram representation of the power spectral density of the function.  Increasing the interval or window size decreases the temporal resolution while decreasing the size of the interval decreases the spectral resolution.  Since the FFTs are not localized in time, the STFT is more suitable for analyzing EEGs, which are transient and unpredictable signals over time.  In Matlab,  'blackman', 'hanning' and 'kaiser' windows can be used as the window parameter for the \code{stft} function.

    The maximum number of levels $L_{max}$ which a signal can be decomposed can be determined by the equation:
    \begin{equation}
        L_{max} = \code{round}(\text{log}_{2}(\frac{N}{N_{w}}-1))
    \end{equation}  
    where $N$ is the length of the signal, $N_{w}$ gives he length of the decomposition filter associated with the chosen mother wavelet and \code{round}, rounds the value to its nearest integer.   

\section{Scalograms and Spectrograms}

\ \ \    Spectrograms are a visual representation of the spectrum of frequencies of a signal as it varies with time.   One can visualize the signal strength at various frequencies present in a particular waveform.  Spectrograms functions result in dB/Hz while in Matlab the power spectrum (\code{spectrum}) of a signal is measured in dB.     If spectrograms are generated by a bank of band-pass filters, by Fourier transform or by wavelet transform in which case they are known as scalograms.  In Matlab, the \code{spectrogram} function uses the short-time Fourier transform (STFT). 

    Various windows (windowing functions) can be used such as Hamming, periodic hann, and Kaiser. Time-frequency representations of biomedical signals can be represented as scalograms. A scalogram is the absolute value of the CWT coefficients of a signal.  Scalograms are a form of transforming CWT into a 3D tensor, with a pre-determined number of layers, in which each layer corresponds to the convolutions of the signal data with a different scale mother wavelet for each layer.  In Matlab, this can be generated using the function \code{cwt}(x,Fs). 

The CWT transforms EEG and ECG signals to the time-frequency domain and a CNN is used to extract features from a 2D scalagram composed of decomposed time-frequency components via multiresolution analysis.  CWT is used for multi-dimensional signal processing and the CNN is used for image feature extraction.  

Spectrograms/scalograms can be used to analyze the energy distribution of the signal. Scalogram images are essentially 2D matrices (time $\times$ frequency) and can be used as images inputs into convolutional neural networks (CNNs) that can be used to perform convolutions to extract meaningful details, features, and information to aid in classifying a signal type. 

 There are several metrics to measure the quality of wavelet denoising.  The signal-to-noise ratio (SNR) is most relevant.  Denote x[n] the original EEG signal and $\hat{x}[n]$ is the denoised EEG signal (which can be obtained by tuning the wavelet parameters via meta-heuristic algorithms), and $N$ is the sampling number:
\begin{equation}
    SNR = 10\text{log}_{10} \bigg [ \frac{\sum_{n=1}^{N} [x(n)]^{2}}{\sum_{n=1}^{N} [x(n) - \hat{x}(n)]^{2}} \bigg]
\end{equation}
    For purposes of deep learning classification, cross entropy will be used as it is the most popular loss function.

\section{ECG Signals}

    \ \ \ An ECG measures heart rhythm activity through electrodes placed on the patient's chest.  An ECG runs on graph paper as waves at 25 mm/sec.  On the time (x-axis), 1 mm square is 0.04 seconds, 5 small squares is 0.2 seconds, and 25 small squares (5 big squares) is 1 second. On the y-axis is voltage.  1 mm square is 0.1 mV and 10 small squares is 1mV.  

    Leads are a pair of electrodes measure the difference in electric potential between either (1) two exploring (active) electrodes attached to the surfae of body (bipolar leads) or (2) one point on th body (exploring) and a virtual reference point (indifferent) electrode with zero electrical potential (unipolar leads).  The wires of the lead make a coplete circuit.  There are 12 leads in an ECG: 
    \begin{itemize}
        \item 3 standard bipolar limb leads (I, II, III) 
        \item 3 augmented unpolar limb leads (aVR, aVL, aVF) 
        \item 6 unipolar (precordial) chest leads (V1, V2, V3, V4, V5, V6)
    \end{itemize}
    Two electrodes are placed on the body (one positive, one negative)  A positive/upwards deflection on the ECG is made by (1) positive changes moving (depolarization) towards a positive electrode or (2) negative charges moving (repolarization) towards a negative electrode.  A negative/downwards deflection on the ECG is made by (1) positive charges moving towards a negative electrode or negative charges moving twoards a positive electrode.  An isoelectric (straight) line is made by no movement of charges (i.e., AV node delay).  

    \ \ \ In an ECG, a complete QRS complex consists of a Q-, R- and S-wave. The QRS complex is the ventricular contraction (systole) consisting of the Q wave, which is the first negative deviation, followed by the R wave, a positive (upward) deviation. Any negative deflection following immediately after the R portion is termed the S wave.  The QRS interval represents the time required for a stimulus to spread through the ventricles (ventricular depolarization) and is normally about $\leq 0.10$ sec (or $\leq$ 0.11 sec when measured by computer)  A deflection is only referred to as a wave if it passes the baseline. If the first wave is negative then it is referred to as Q-wave. If the first wave is not negative, then the QRS complex does not possess a Q-wave, regardless of the appearance of the QRS complex.  All positive waves are referred to as R-waves. The first positive wave is simply an \enquote{R-wave} (R). The second positive wave is called \enquote{R-prime wave} (R’). If a third positive wave occurs (rare) it is referred to as \enquote{R-bis wave} (R”).  Any negative wave occurring after a positive wave is an S-wave.

    Large waves are referred to by their capital letters (Q, R, S), and small waves are referred to by their lower-case letters (q, r, s). However, all three waves may not be visible and there is always variation between the leads. Some leads may display all waves, whereas others might only display one of the waves. Regardless of which waves are visible, the wave(s) that reflect ventricular depolarization is always referred to as the QRS complex.  The QRS complex is measured from beginning of the Q wave until the end of the S wave.
    
    The R wave reflects depolarization of the main mass of the ventricles –hence it is the largest wave. the S wave signifies the final depolarization of the ventricles, at the base of the heart.  Poor R wave progression in right precordial leads is a relatively common electrocardiogram (ECG) finding that indicates possible prior anterior myocardial infarction (MI); however, it is observed frequently in apparently normal individuals. In contrast, reversed R wave progression (RRWP) may be more specific to cardiac disorders.  In particular, though RRWP is rare in daily clinical practice; it is a highly indicative marker for cardiac disease, particularly ischemtic heart disease (IHD), or coronary artery disease, with left anterior descending (LAD) artery stenosis.

    There are ECG wave types including P-waves which are caused by atrial depolarization and occur from the time of electrical impulse from sinoatrial (SA) node and precedes atrial contraction by 0.02 seconds.   T-waves are caused by ventricular repolarization and occur during the latter part of systole.  A normal TP segment is calculated from the end of a T-wave to the beginning of a P-wave.  The time interval of TP is from the end of ventricular repolarization until the next atrial depolarization.  It represents ventricular filling.

\begin{figure}[H]
    \centering \includegraphics[width=0.8\columnwidth]{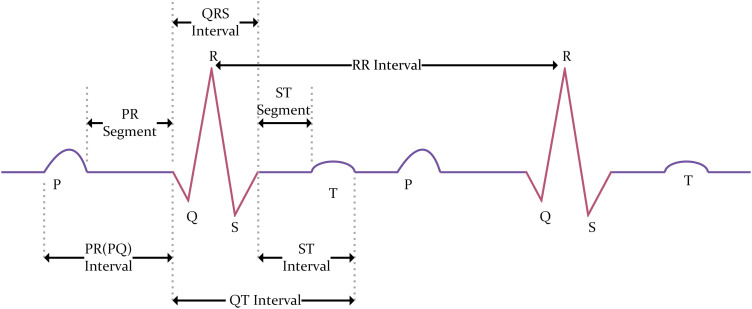} 
    \caption{ECG Signal Diagram. \cite{Almousa:2023}} 
    \label{fig:ecg2}
\end{figure}

\begin{figure}[H]
    \centering \includegraphics[width=0.4\columnwidth]{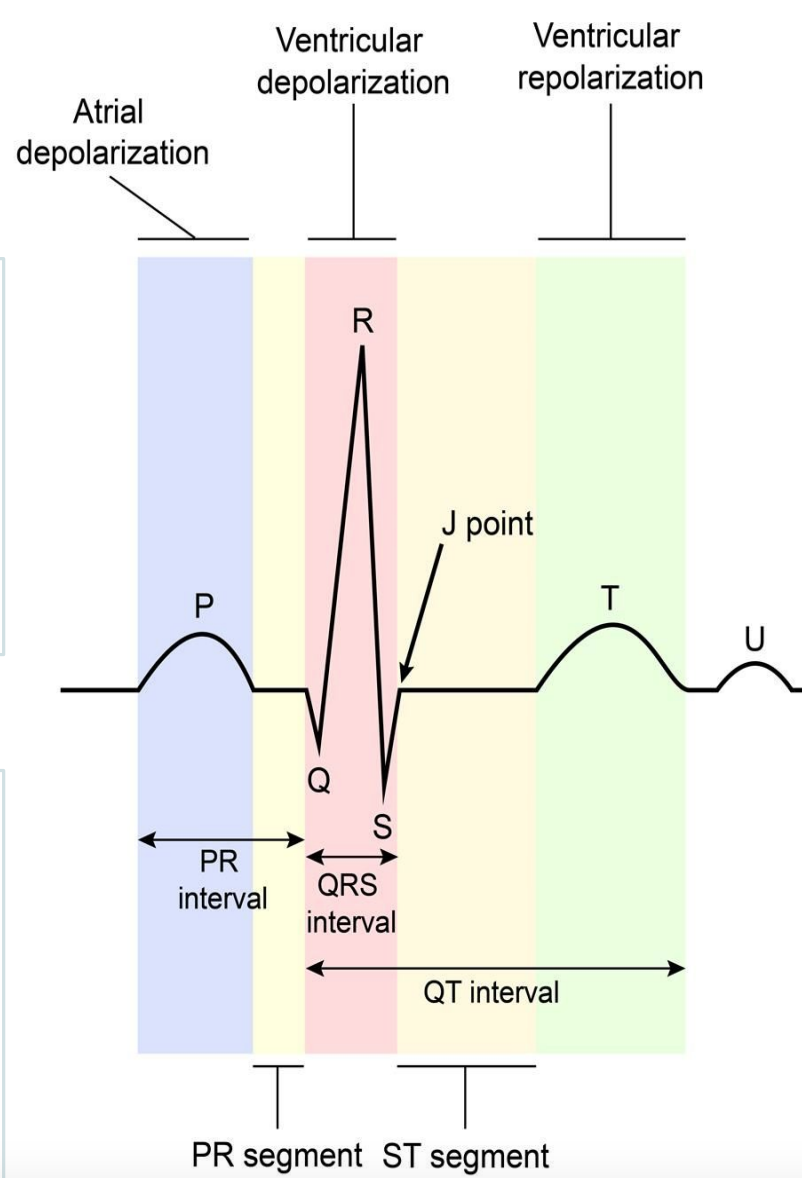} 
    \caption{QRS Complex: Repolarization. \cite{Dominguez:2019}} 
    \label{fig:QRS1}
\end{figure}

    Figure \ref{fig:nodes} illustrates various combinations of Q, R, and S wave signals.
    \begin{figure}[H]
    \centering    \includegraphics[width=0.75\columnwidth]{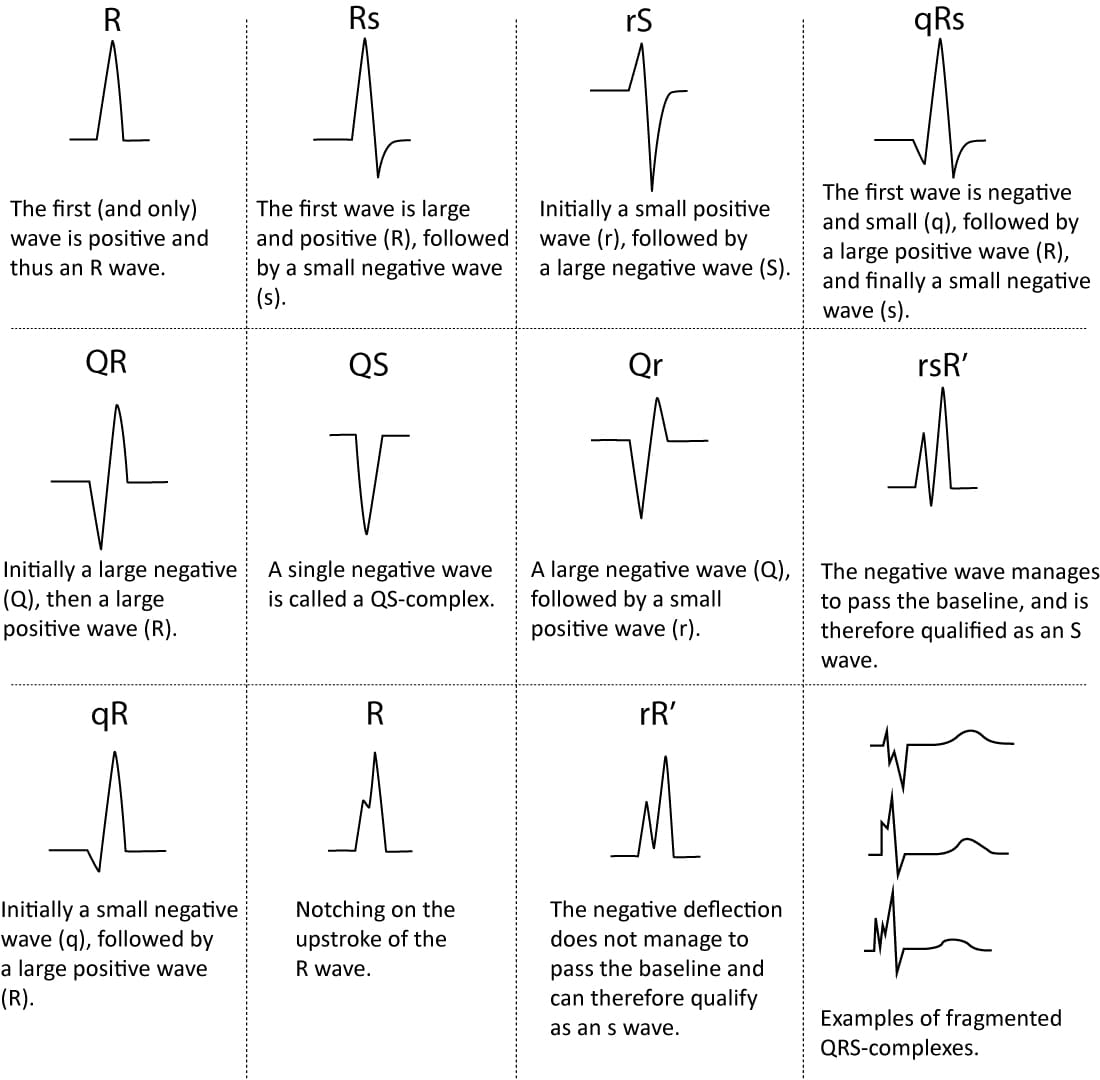} 
    \caption{Various QRS signals. \cite{ecgwaves:2023}} 
    \label{fig:nodes}
\end{figure} 
Depolarization of the ventricles generate three large vectors, which explains why the QRS complex is composed of three waves. Figure \ref{fig:nodes} illustrates the vectors in the horizontal plane.  Depolarization means positive charges are flowing int cells such as the depolarization of one cell stimulates the depolarization of adjacent cells through gap junctions.  Repolarization means negative charges are flowing in.
\begin{figure}[H]
    \centering \includegraphics[width=0.75\columnwidth]{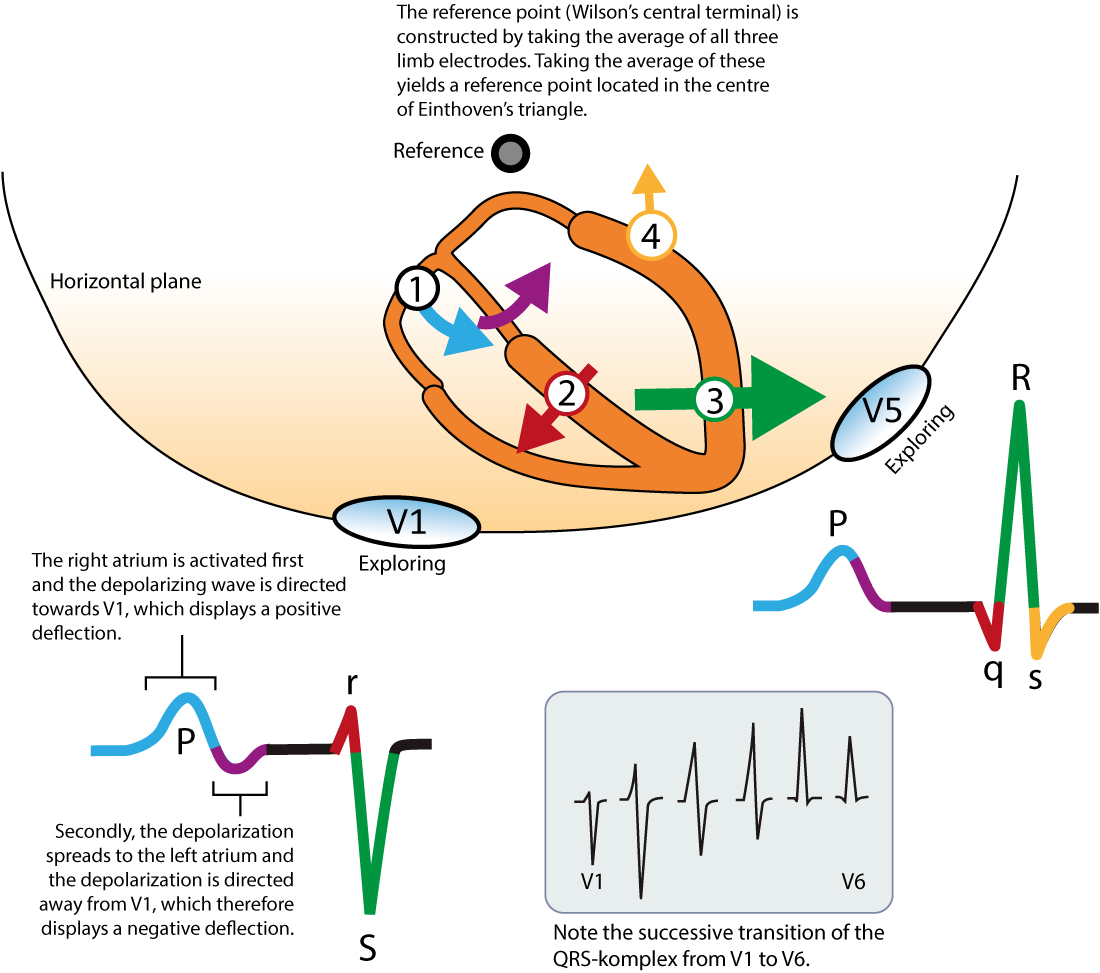} 
    \caption{Various QRS signals. \cite{ecgwaves:2023}} 
    \label{fig:nodes}
\end{figure}
    Figure \ref{fig:ecg2} shows a diagram of a typical ECG signal cycle.  The signal consists of three parts: the P wave, the QRS complex, and the T wave.   The diagram shows critical parameters like the PQ, ST, QRS, and QT intervals.
    Figure \ref{fig:QRS1} shows breakdown of the ECG signal and heartbeat cycle of atrial depolarization, ventricular polarization, and and ventricular repolarization. 

\begin{figure}[H]
    \centering \includegraphics[width=0.75\columnwidth]{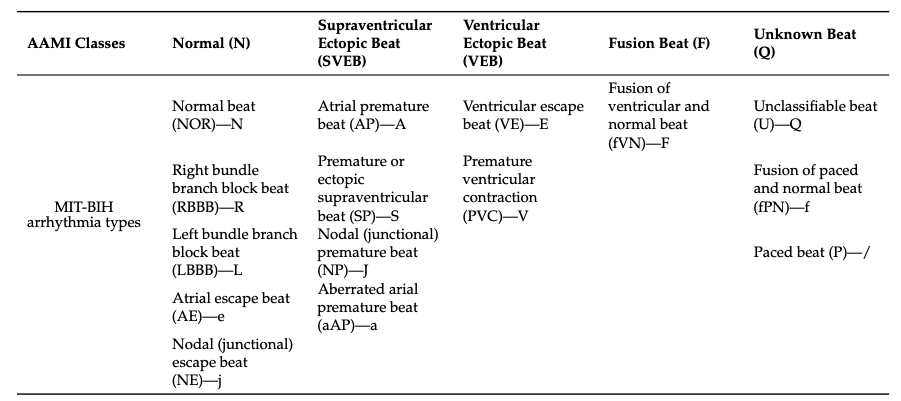} 
    \caption{Types of Cardiac Arrhythmia Signals}
    \label{fig:beats}
\end{figure}

\section{ECG Signal Processing}

\ \ \ In ECG signal processing, there are five main categories of arrhythmia beats.  Figure \ref{fig:beats} shows a detailed breakdown of the ECG signal classes:
\begin{itemize}
\item N : Non-ecotic beats (normal beats)
\item S: Supraventricular ectopic beats
\item V: Ventricular ectopic beats
\item F: Fusion beats
\item Q: Unknown beats
\end{itemize}
    ECG classification has various challenges including aliasing.  The ECG signal is \enquote{composed of different frequency components and noises, which increases the difficulty of deep learning-based methods to extract discriminant feature \cite{Wang:2021}.} 

    Thus, noise removal is important in EEG and ECG signal processing.  The presence of different artifacts such as ocular artifacts (OA) and electromyograms must be considered. Normally EEG signals falls in the frequency range of DC to 60 Hz and amplitude of 1-5 $\mu$v. Ocular artifacts have similar statistical properties of EEG signals, often interfere with EEG signal, thereby making the analysis of EEG signals more complex.  On a 12 lead ECG this is usually a 10 second recording from Lead II. Usually for ECG signals, the frequency range over 80Hz is noise.  There is also power line noise between 50-60 Hz.  A notch filter can be used to remove these noises.  

    Using a db6 wavelet transform, we can use a 1D wavelet decomposition using the Matlab \code{mdwtdec} function as shown in the Appendix.
\subsection{ECG Conversion from 1D Signal to 2D Image}
\begin{figure}[H]
    \centering \includegraphics[width=0.9\columnwidth]{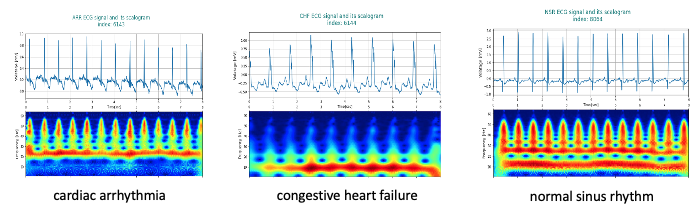} 
    \caption{Scalograms for Different ECG Signal Types} 
    \label{fig:scalogram1}
\end{figure}

\begin{figure}[H]
    \centering \includegraphics[width=0.75\columnwidth]{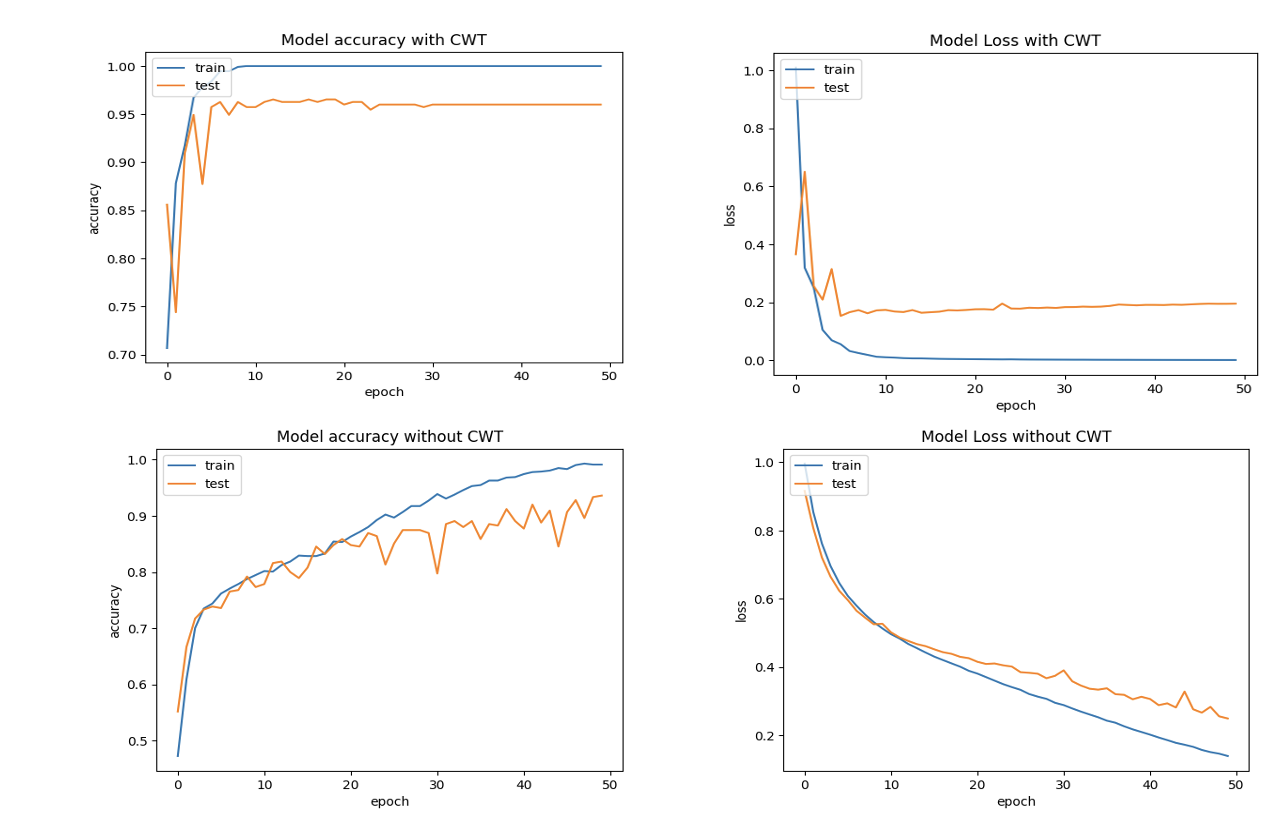} 
    \caption{Loss and Accuracy Plot With and Without Denoising} 
    \label{fig:lossplot}
\end{figure}

Various scalograms using wavelet transforms generated for different types of ECG signals including cardiac arrhythmia (ARR), congestive heart failure (CHF), and normal sinus rhythm (NSR) are illustrated in Figure \ref{fig:scalogram1}
    As shown in Figure \ref{fig:lossplot}, using the coefficients from the wavelet decomposition as the input into the CNN generates quick loss convergence and high accuracy compared to deep learning without wavelet denoise preprocessing.
    \ \ \ A signal's R-peak placements are determined by its highest amplitude \cite{Rizwan:2022}. Ventricular depolarization cannot be detected during the refractory period, which is 200 ms.  The Pan-Tompkins Algorithm \cite{Pan:1985} can used to detect R waves from the QRS complex present in ECG signals to measure the heart rate of a patient.  The algorithm works by analyzing the slope, amplitude and width of the QRS complexes present in the filtered ECG signal. The ECG signal is filtered so as to reduce noise and decrease detection thresholds, thereby increasing the sensitivity towards detection of the QRS complex.

\begin{figure}[H]
    \centering \includegraphics[width=0.9\columnwidth]{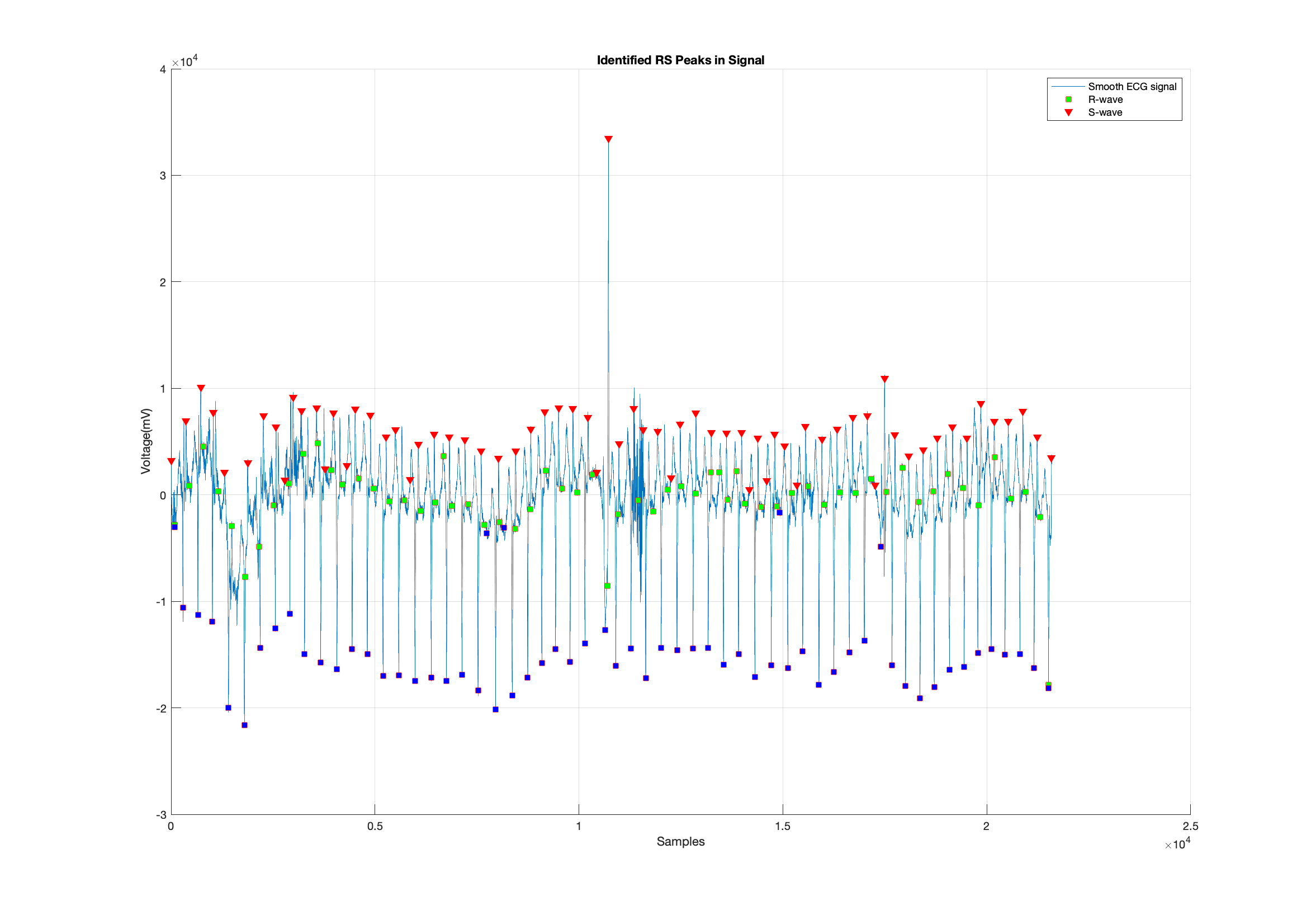} 
    \caption{Detection of peaks of R waves using Pan-Tompkins.  \cite{Abbas:2017}} 
    \label{fig:peak4}
\end{figure}

\begin{figure}[H]
    \centering \includegraphics[width=0.7\columnwidth]{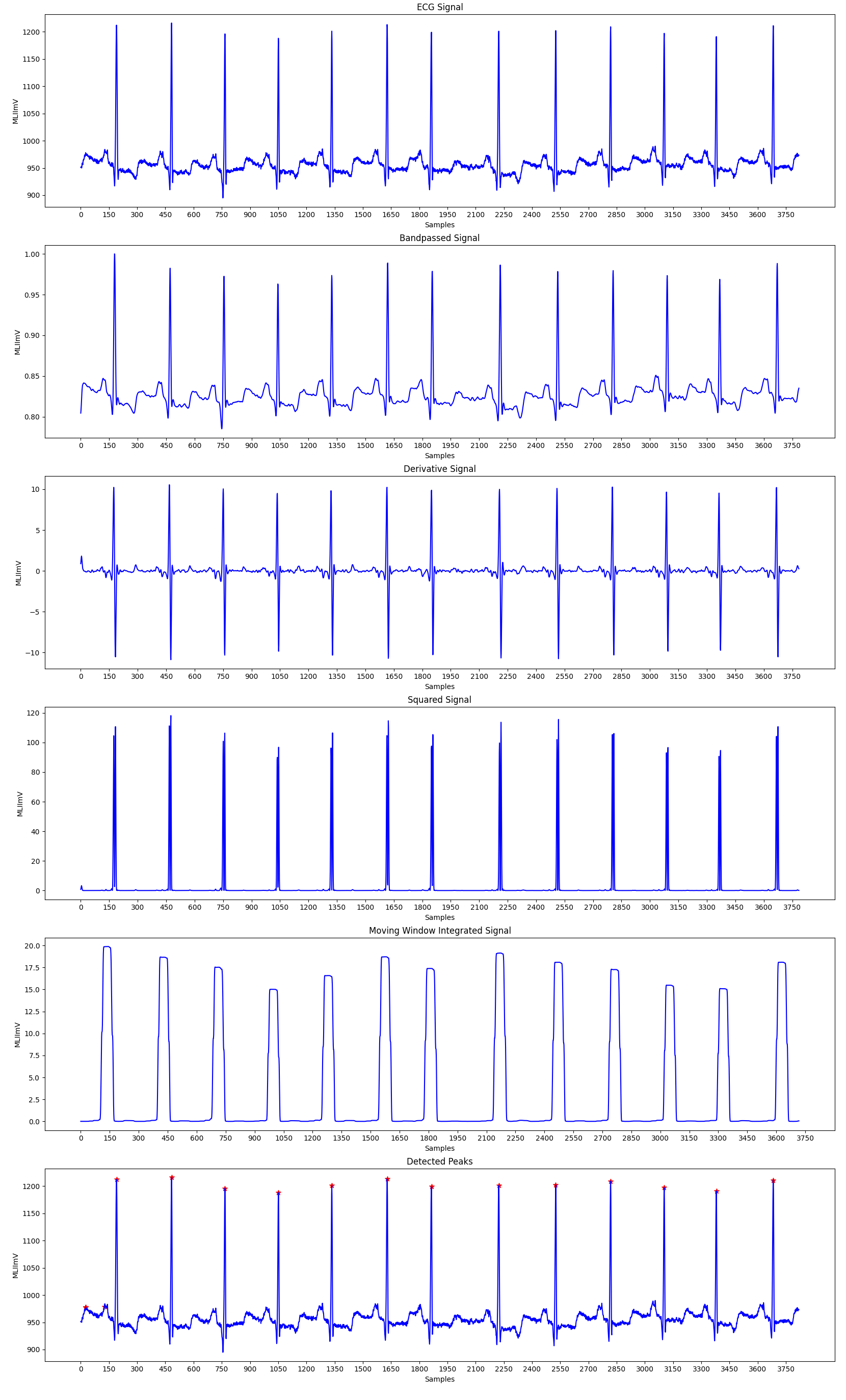} 
    \caption{Derivative ECG Signals} 
    \label{fig:derivative}
\end{figure}

\begin{figure}[H]
    \centering \includegraphics[width=0.7\columnwidth]{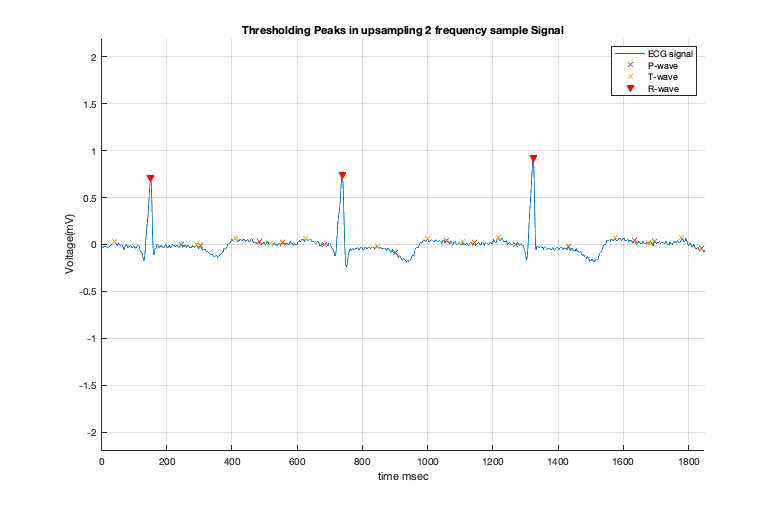} 
    \caption{Thresholding Peaks of P,T, and S-waves by Upsampling by 2. \cite{Lim:2018}} 
    \label{fig:peak3}
\end{figure}

    The algorithm can be divided into various phases, the first phase consists of applying the filtered on the input ECG signal, followed by peak detection in the filtered signal. The peak detection again works in three phases: Learning Phase 1, Learning Phase 2 and Detection.   Figure \ref{fig:peak4} illustrates peak detection of R-waves using the Pan-Tompkins algorithm.
    Learning Phase 1 is required to initialize the signal and noise thresholds followed by Learning Phase 2 in which the RR intervals and the RR limit values are initialized. The detection phase works by adjusting the thresholds appropriately and recognizing the QRS complexes. A dual threshold is used to increase the detection sensitivity along with the improvement in the signal to noise ratio by the bandpass filter \cite{Sharma:2021}. Implementation details are provided in \cite{Abbas:2017}, \cite{masud:2018}, \cite{Lim:2018}.
    Figure \ref{fig:derivative} illustrates various signals that can be generated from an ECG (or any) signal: bandpassed signal, derivative signal, squared signal, moving average window integrated signal, and detected peaks signal. 
    Figure \ref{fig:peak3} illustrates using thresholding to find peaks of P,T, and S-waves of a sample ECG signal by upsampling the frequency by 2. 

\section{ECG Classification} 

\ \ \ Classifiers seeks to classify heartbeat signals into five superclasses: normal beat (N), supraventriculalar ectopic beat (SVEB), ventricular ectopic beat (VEB), fusion beat (F), and unknown beat (Q).  Different algorithms have been used, including artificial neural networks (ANNs), support vector machines (SVMs), multi-view-based learning, and linear discriminants (LDs) for classification \cite{Xu:2018, Wang:2021}.  

    However, \enquote{despite the good performance achieved by these methods, the ECG waves and their morphological characteristics of different patients have significant variations, and for the same patient, the ECG waves at different times are also different. The fixed features used in these methods are not [sufficient] to accurately distinguish arrhythmia of different patients. Recently, with the rapid development of deep neural networks, deep learning-based methods have attracted more and more attention. Deep learning, as a representation learning method, can automatically extract discriminant features from the training data \cite{Wang:2021}.}

    The MIT-BIH Arrhythmia Database is the benchmark for deep learning classification of ECG signals.  Using this database, Ahmad, et. al. \cite{Ahmad:2021} introduce two computationally efficient multimodal fusion frameworks for ECG heart beat classification called Multimodal Image Fusion (MIF) and Multimodal Feature Fusion (MFF). At the input of these frameworks, the raw ECG data is converted into three different images using Gramian Angular Field (GAF), Recurrence Plot (RP) and Markov Transition Field (MTF). They perform experiments on PhysioNet’s MIT-BIH dataset for five distinct conditions of arrhythmias which are consistent with the AAMI EC57 protocols and on PTB diagnostics dataset for Myocardial Infarction (MI) classification.

\subsection{Experiment 2: MIT-BIH Arrhythmia Database}
    \ \ \ The MIT-BIH Arrhythmia Database contains 48 half-hour excerpts of two-channel ambulatory ECG recordings, obtained from 47 subjects studied by the BIH Arrhythmia Laboratory between 1975 an 1979.\footnote{\url{https://www.physionet.org/content/mitdb/1.0.0/}}   23 recordings were chosen at random from a set of 4000 24-hour ambulatory ECG recordings collected from a mixed population of inpatients (about $60\%$) and outpatients (about $40\%$) at Boston's Beth Israel Hospital; the remaining 25 recordings were selected from the same set to include less common but clinically significant arrhythmias that would not be well-represented in a small random sample.

    The recordings were digitized at 360 samples per second per channel with 11-bit resolution over a 10 mV range.  Two or more cardiologists independently annotated each record; disagreements were resolved to obtain the computer-readable reference annotations for each beat (approximately 110,000 annotations in all) included with the database.

\begin{figure}[H]
    \centering \includegraphics[width=0.8\columnwidth]{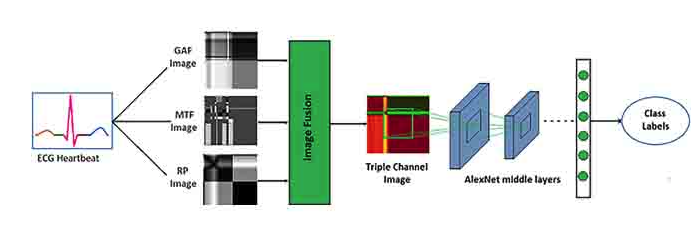} 
    \caption{Multi-modal fusion model. \cite{Ahmad:2021}} 
    \label{fig:gaf}
\end{figure}

    Ahmad, et. al. \cite{Ahmad:2021} introduced a multi-modal image fusion (MIF) and multi-modal feature (FF) fusion model that use combine (fuse) Gramian angular fusion (GAF), recurrence plot (RP), and Markov Transition Field (MTF) together to create a triple channel image that gets input into the CNN as illustrated in Figure \ref{fig:gaf}.  Wang et. al \cite{Wang:2015} proposed the GAF algorithm that enables 1D signals to be converted to 2D images.  

    Consider a time series 
    $X = \{x_{1},x_{2},\dots,x_{N}\}$ with $N$ observations.  Normalize $X$ so that all values can be in the range of [-1,1] or [0,1] which can be expressed, respectively, as:
    \begin{align}
        \tilde{x}^{i}_{-1} &= \frac{(x_{i} - \text{max}(X)) + (x_{i} - \text{min}(X))}{\text{max}(X) - \text{min}(X)} \\
        \tilde{x}^{i}_{0} &= \frac{(x_{i} - \text{min}(X)}{\text{max}(X) - \text{min}(X)}
    \end{align} 
    Convert the 1D time series from Cartesian coordinates to polar coordinates: \begin{equation}
    \begin{cases}
        \phi_{i} = \arccos(\tilde{x}_{i}), \ \ \ -1 \leq \tilde{x}_{i} \leq 1, \tilde{x}_{i} \in \tilde{X} \\
        r_{i} = \frac{i}{N}, \ \ \ \  i \in N \nonumber    
    \end{cases} 
    \end{equation}
    where the inverse cosine of the normalize observation $\tilde{x}_{i}$ is taken as the angle $\phi_{i}$ in the polar coordinate system , and the time label $i/N$ is taken as the radius.   The angle range of the cosine function corresponding to the dates within the range [0,1] is $[0,\pi/2]$ and the angle corresponding to the data in the range of [-1,1] is $[0,\pi]$ \cite{Xu:2020}.
    \subsection{Grammian Angular Field}
    \ \ \ Define the inner product $\langle x, y \rangle = x \cdot y - \sqrt{1-{x}^2}\cdot \sqrt{1-y^{2}}$, \textbf{G} is a Grammian matrix:
    \begin{equation}
        \begin{bmatrix}
            \langle \tilde{x}_{1}, \tilde{x}_{1} \rangle & \cdots &\langle \tilde{x}_{1}, \tilde{x}_{n} \rangle \\
            \langle \tilde{x}_{2}, \tilde{x}_{1} \rangle & \cdots & \langle
            \tilde{x}_{2}, \tilde{x}_{n} \rangle \\
            \vdots & \ddots & \vdots \\
            \langle \tilde{x}_{n}, \tilde{x}_{1} \rangle & \cdots & 
            \tilde{x}_{n}, \tilde{x}_{n} \rangle 
        \end{bmatrix}
    \end{equation}
    The Gramian angular field (GAF) preserves temporal dependencies.  As time increases, the position moves from top-left to bottom-right in \textbf{G}.

    By using a polar coordinate representation of the data, then as time elapses, the sequence value varies from the original amplitude change to the angular change in the polar coordinate system.  By \enquote{calculating the sum/difference of the trigonometric function among sampling point, the time correlation among them is identified from the perspective of angle.  Gramian angular summation field (GASF) and Gramian angular difference field (GADF) are defined as follows, respectively \cite{Xu:2020}}: 
    \begin{align}
        GASF = \begin{bmatrix}
                    \text{cos}(\phi_{1} + \phi_{1}) & \cdots & \text{cos}(\phi_{1} + \phi_{n}) \\
                    \text{cos}(\phi_{2} + \phi_{1}) & \cdots & \text{cos}(\phi_{2} + \phi_{n}) \\
                    \vdots & \ddots & \vdots \\
                    \text{cos}(\phi_{n} + \phi_{1}) & \cdots & \text{cos}(\phi_{n} + \phi_{n}) \nonumber 
               \end{bmatrix} \\
        GADF = \begin{bmatrix}
                    \text{sin}(\phi_{1} - \phi_{1}) & \cdots & \text{sin}(\phi_{1} - \phi_{n}) \\
                    \text{sin}(\phi_{2} - \phi_{1}) & \cdots & \text{sin}(\phi_{2} - \phi_{n}) \\
                    \vdots & \ddots & \vdots \\
                    \text{sin}(\phi_{n} - \phi_{1}) & \cdots & \text{sin}(\phi_{n} - \phi_{n}) \nonumber 
               \end{bmatrix}
    \end{align}
    
    The GAF algorithm is adopted to transform the one-dimensional time series into the two-dimensional images through three steps of scaling, coordinate axis transformation and trigonometric function, so as to apply the computer vision technology to the study of time \cite{Xu:2020}.  Thus, through the polar coordinates, GAFs represent the mutual correlations between each pair of points/phases by the superposition of nonlinear cosine functions \cite{Wang2:2015}.

    \subsection{Recurrence Plot}
\begin{figure}[H]
    \centering \includegraphics[width=0.6\columnwidth]{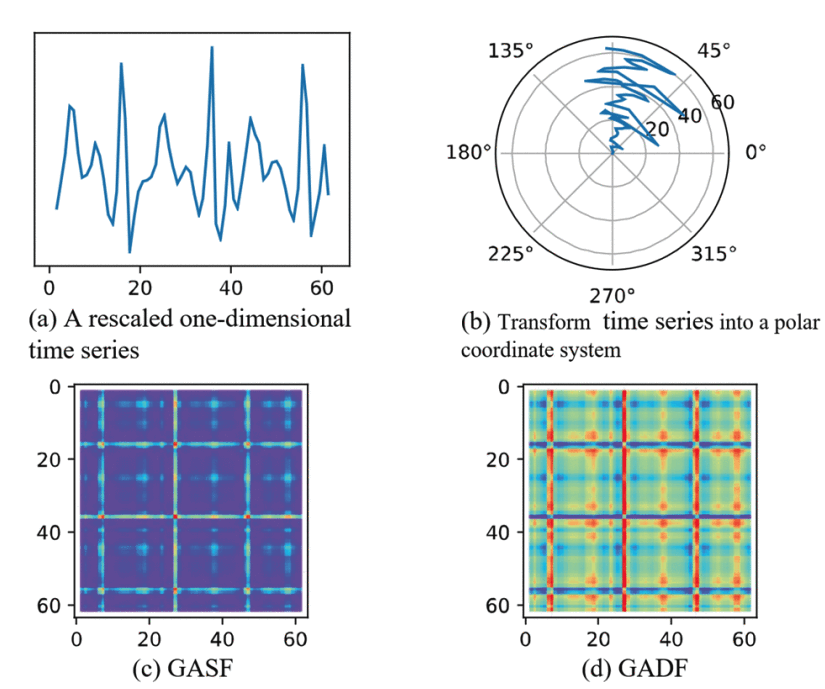} 
    \caption{1D time series converted to 2D images in polar coordinates. \cite{Xu:2020}} 
    \label{fig:polar} 
\end{figure}

    \ \ \ A recurrence plot (RP) is a fundamental property of a dynamic system such as electrical signals generated by the human heart, which is difficult to detect in serial time-domain signals \cite{Zhang:2022}.   The RP approach was introduced to explore the phase space trajectory in a higher-dimensional space and to show the recurrent behavior of the time series.  An RP can be formulated as:
    \begin{equation}
        \textbf{R}_{i,j} = \mathbb{H}(\mathbf{\epsilon} - \lvert \rvert \textbf{x}_{i} - \textbf{x}_{j} \lVert \rvert) \ \ \ \ i,j = 1,\dots,N
    \end{equation}  
    where $N$ is the number of time series $\mathbf{x}_{i}$, $\epsilon$ is predefined distance, $\lvert \rvert \cdot \lvert \rvert$ is an L2 norm, $\mathbb{H}$ is the Heaviside function where $\mathbb{H} = 0$ if $z < 0$ and 1 otherwise.  The R-matrix captures pairwise distances between time series data points. In an un-threshold approach, the R-matrix can be defined as 
    \begin{equation}
        \textbf{R}_{i,j} = \lvert \rvert \textbf{x}_{i} - \textbf{x}_{j} \lvert \rvert \ \ \ \ i,j = 1,\dots,N
    \end{equation}    
    Figure \ref{fig:polar} shows the mapping relationship between one-dimensional time series and two-dimensional images. The time series is transformed into a polar coordinate system according to (3). The GASF and GADF images can be obtained by equations (2.14) and (2.14), respectively.
\subsection{Markov Transition Field}
\begin{figure}[H]
    \centering \includegraphics[width=0.75\columnwidth]{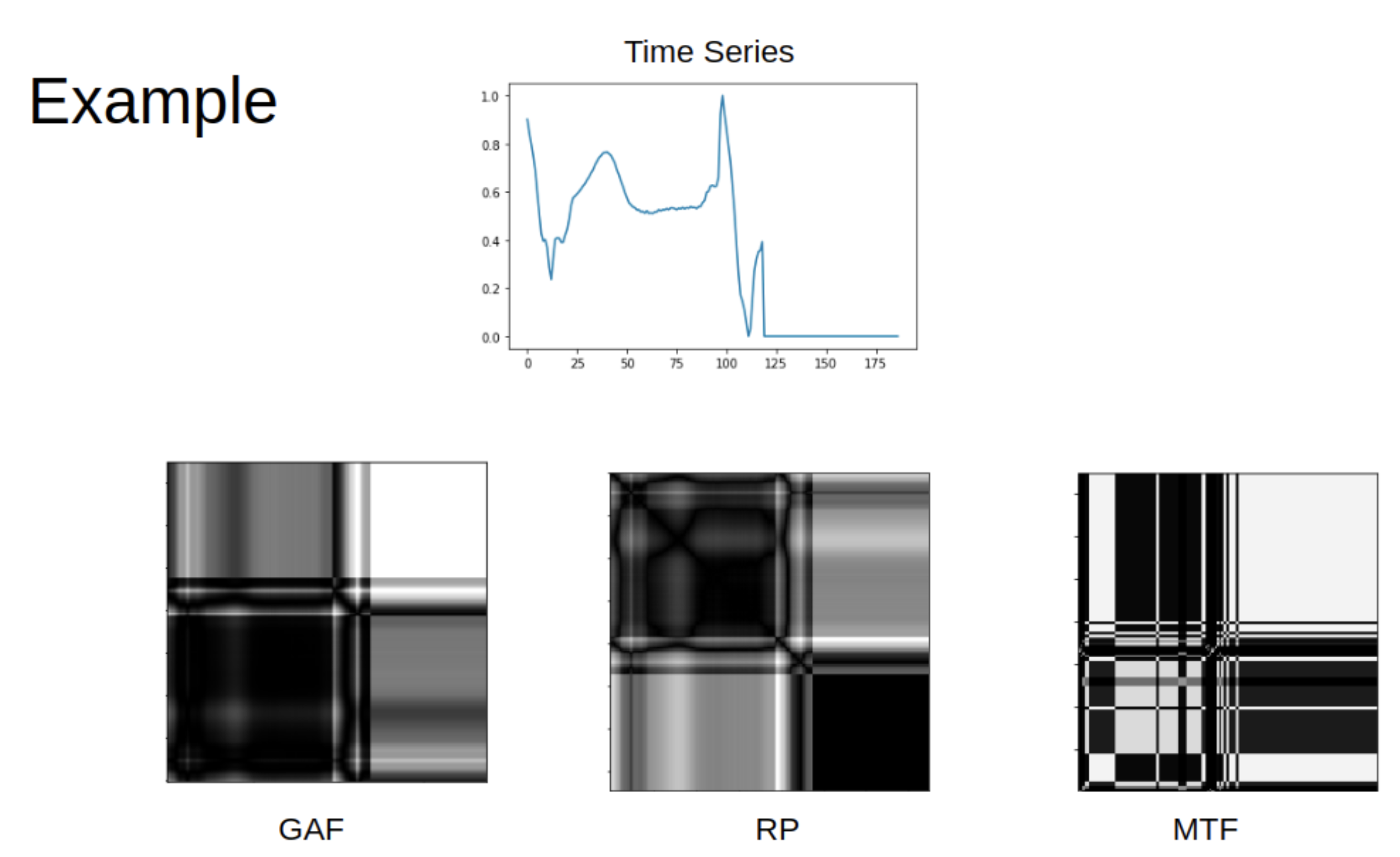} 
    \caption{Example of GAF, RP, and MTF images. \cite{Ahmad:2021}} 
    \label{fig:gaf2}
\end{figure}

\begin{figure}[H]
    \centering \includegraphics[width=0.6\columnwidth]{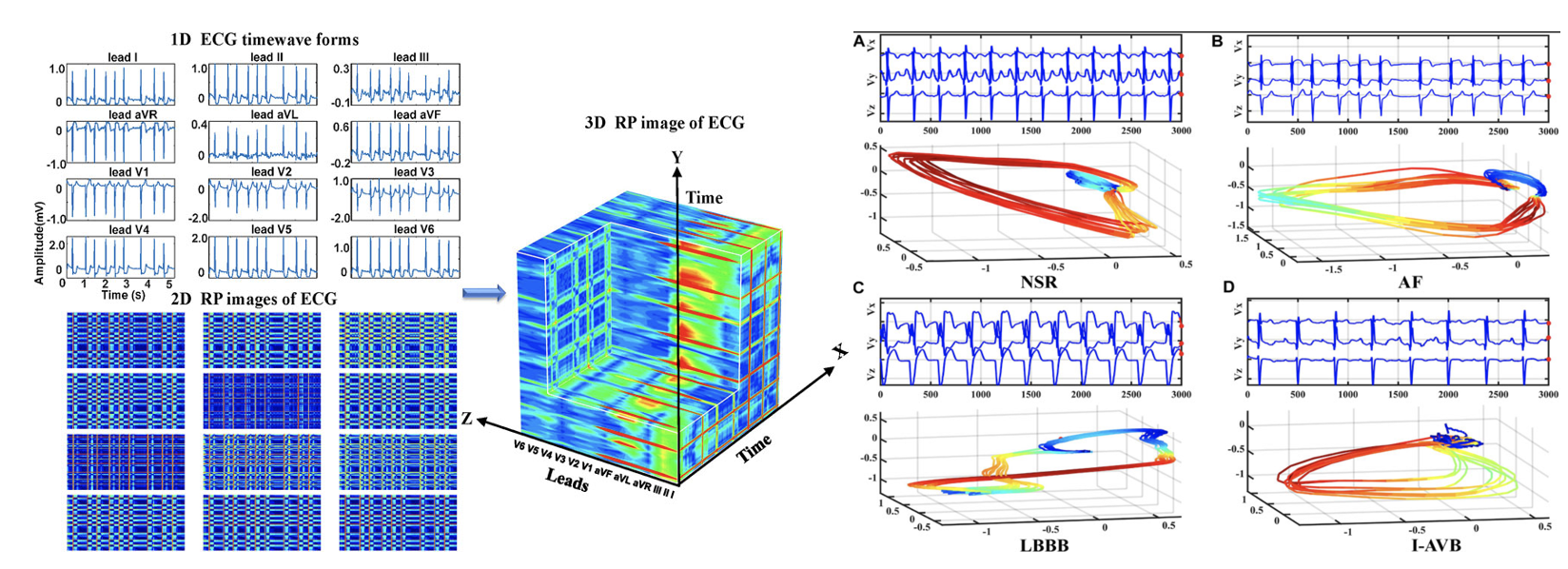} 
    \caption{Illustration of 3D ECG Signal Recurrence Plots. \cite{Zhang:2022}} 
    \label{fig:3D}
\end{figure}

    \ \ \ Markov transition probabilities sequentially can preserve information in temporal dimension.   For a given time series, one finds its $Q$ quantile bins and assigns each $x_{i}$ to its corresponding bin $q_{j} (j \in [1,Q])$.   Thus, one constructs a $Q \times Q$ weighted adjacency matrix $W$ by counting transitions among quantile bins in the manner of a first-order Markov chain along the time axis.   Denote $w_{i,j}$ as the frequency with which a point in quantile $q_{j}$ is followed by a point in quantile $q_{i}$  Define the Markov Transition Field (MTF) as follows:
    \begin{equation}
        M = \begin{bmatrix}
                v_{i,j|x_{1} \in q_{i}, x_{1} \in q_{j}} & \cdots & v_{i,j|x_{1} \in q_{i}, x_{n} \in q_{j}} \\
                v_{i,j|x_{2} \in q_{i}, x_{1} \in q_{j}} & \cdots & v_{i,j|x_{2} \in q_{i}, x_{n} \in q_{j}} \\
                \vdots & \ddots & \vdots \\
                v_{i,j|x_{n} \in q_{i}, x_{1} \in q_{j}} & \cdots & v_{i,j|x_{n} \in q_{i}, x_{n} \in q_{j}}
            \end{bmatrix}
    \end{equation}
    \ \ \ One builds a $Q \times Q$ Markov transition matrix $W$ by dividing the data (magnitude) into $Q$ quantile bins.  The quantile bins that contain the data at time steps $i$ and $j$ (temporal axis) are $q_{i}$ and $q_{j}$ ($q \in [1,Q]$).   $M_{i,j}$ in MTF denotes the transition probability of $q_{i} \rightarrow q_{j}$ \cite{Wang2:2015}  Thus, the transition probabilities along the magnitude axis are spread into the MTF matrix by considering temporal positions.  From the patient ECG signals, GAF, RP, and MTF images are generated as illustrated in Fig \ref{fig:gaf2}  They are then input into the CNN where they are fused as illustrated in Figure \ref{fig:gaf2}
    Using multi-modal image (MIF) fusion \cite{Ahmad:2021} with a a ResNet CNN, the confusion matrix generated for the ECG classes N, S, V, F, and Q are:
\begin{figure}[H]
    \centering \includegraphics[width=0.6\columnwidth]{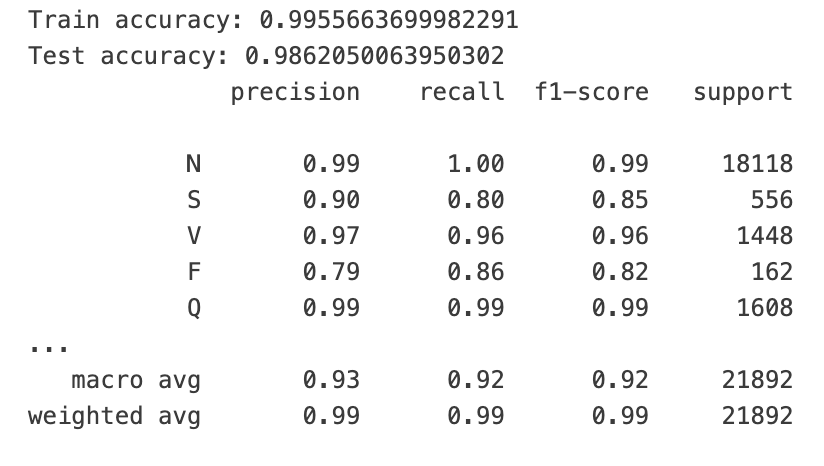} 
    \caption{Confusion matrix using multi-modal image fusion. Source: \cite{Tabassum:2020}} 
    \label{fig:ecg_accuracy}
\end{figure}
    Zhang, et. al (2022) \cite{Zhang:2022} demonstrated converting 1D and 2D ECG images into 3D recurrence plot images as illustrated in Figure \ref{fig:3D} an lead to better classification than just 2D images.
\section{EEG Signals}
\begin{figure}[H]
    \centering \includegraphics[width=0.7\columnwidth]{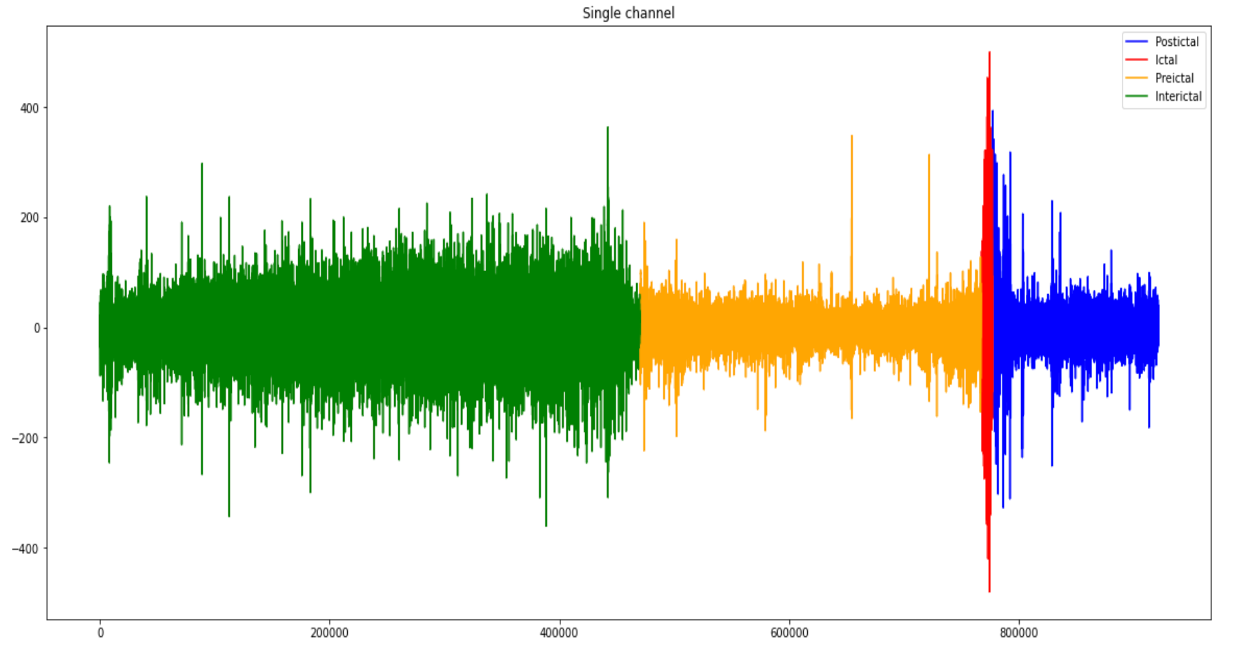} 
    \caption{Seizure states: interictal, preictal, ictal, and post-ictal} 
    \label{fig:ictal} 
\end{figure}

\begin{figure}[H]
    \centering \includegraphics[width=0.7\columnwidth]{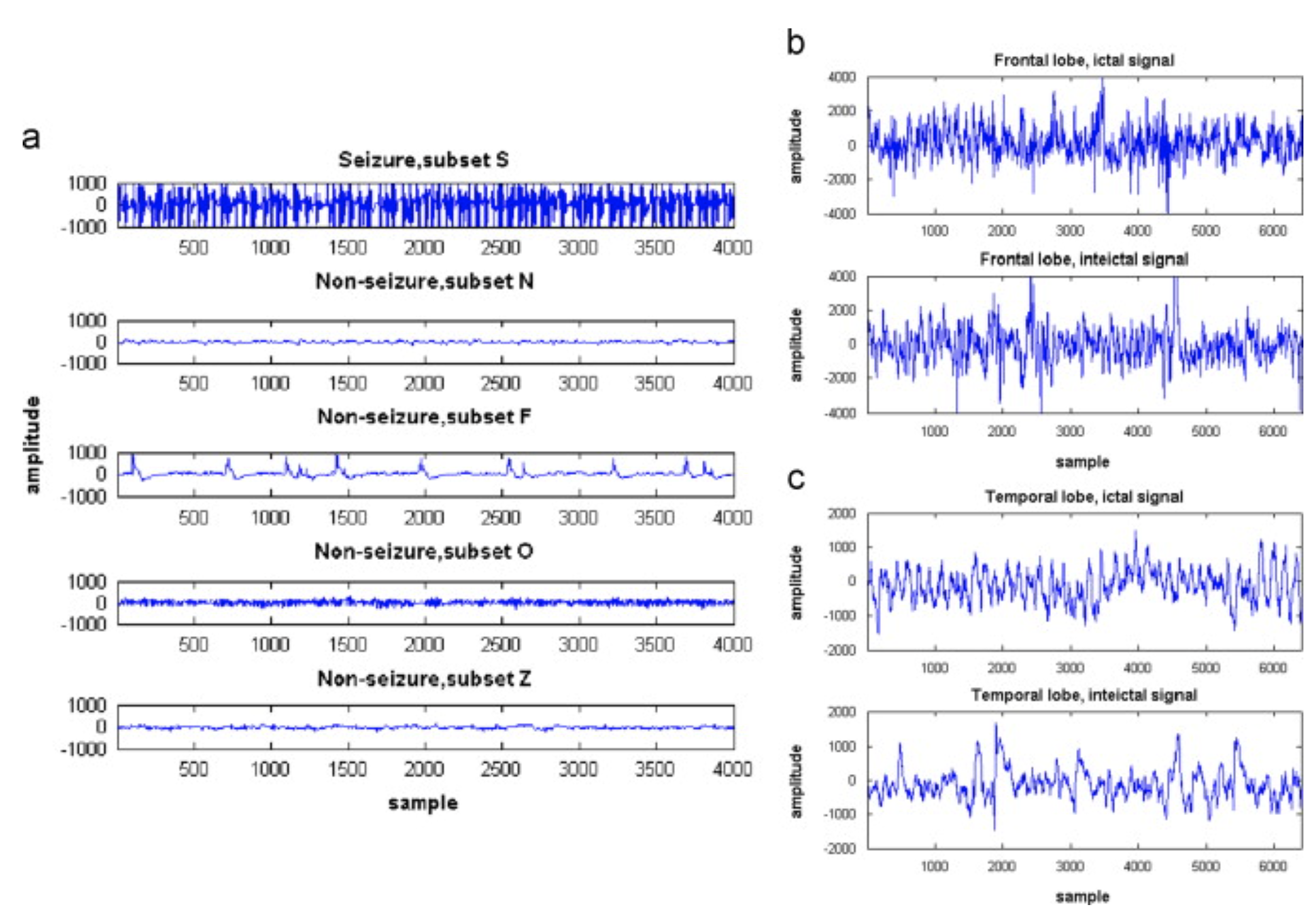} 
    \caption{Illustration of amplitude of ictal and inteictal signals.  \cite{Sirjani:2022}}  
    \label{fig:ictal2} 
\end{figure}

\ \ \ There is emerging evidence that the temporal dynamics of brain activity can be classified into 4 states: (1) interictal (between seizures, or baseline), (2) preictal (prior to seizure), (3) ictal (seizure), and (4) post-ictal (after seizures). Seizure forecasting requires the ability to reliably identify a preictal state that can be differentiated from the interictal, ictal, and postictal states as illustrated in Figure \ref{fig:ictal} \cite{rwave:2023}.
    Figure \ref{fig:ictal2} illustrates the amplitude of ictal and inteictal signals taken from the front and temporal lobe.
The primary challenge in seizure forecasting is differentiating between the preictal and interictal states. Noise removal is important in EEG and ECG signal processing.  The presence of different artifacts such as ocular artifacts (OA) and electromyograms (measure the electrical activity of muscles at rest and at contraction) must be considered. Normally EEG signals falls in the frequency range of DC to 60 Hz and amplitude of 1-5 $\mu$v. Ocular artifacts have similar statistical properties of EEG signals, often interfere with EEG signal, thereby making the analysis of EEG signals more complex.

\begin{figure}[H]
    \centering \includegraphics[width=0.75\columnwidth]{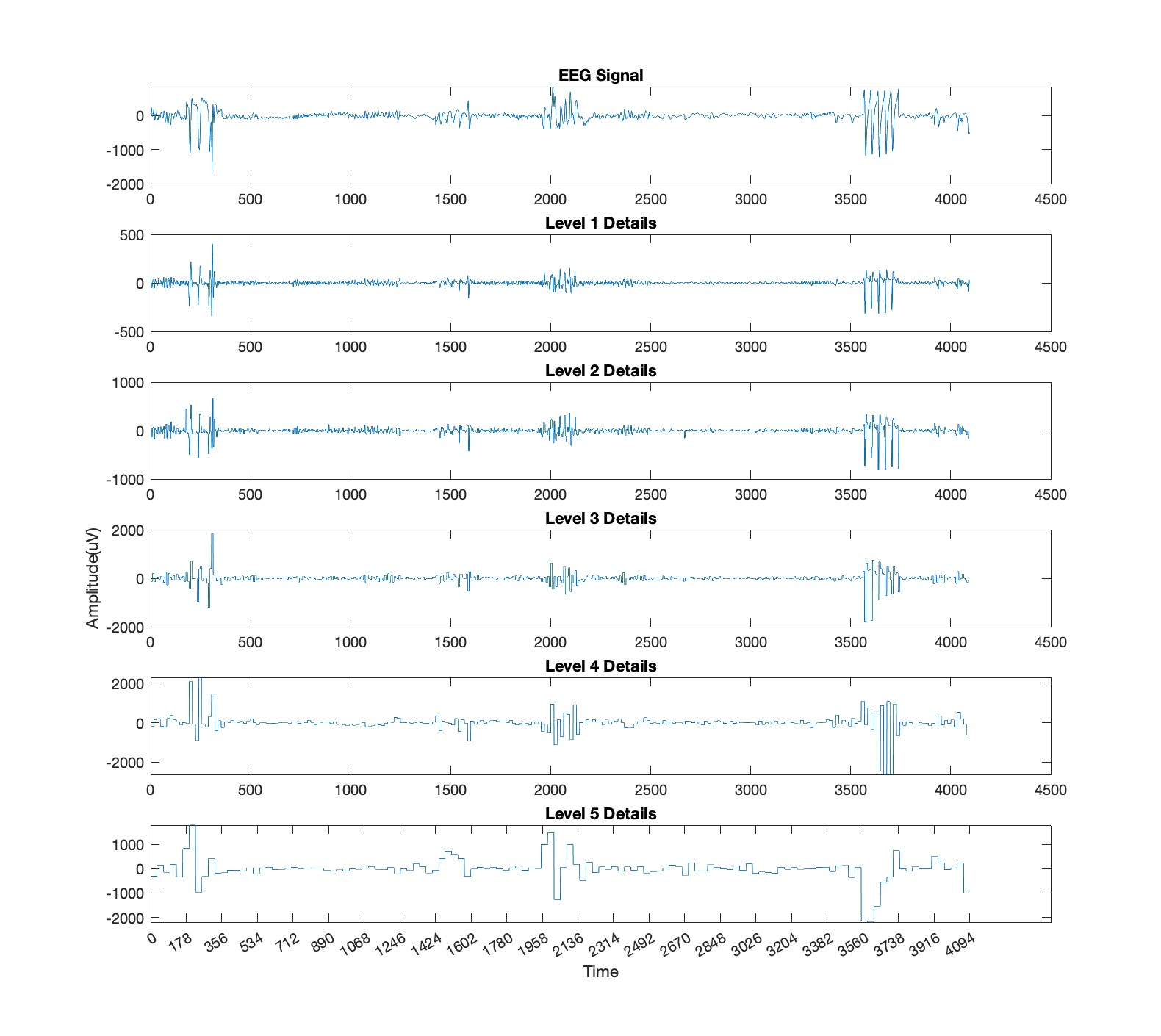} 
    \caption{EEG Signal Wavelet Decomposition} 
    \label{fig:waveletdec}
\end{figure}

\begin{figure}[H]
    \centering \includegraphics[width=0.75\columnwidth]{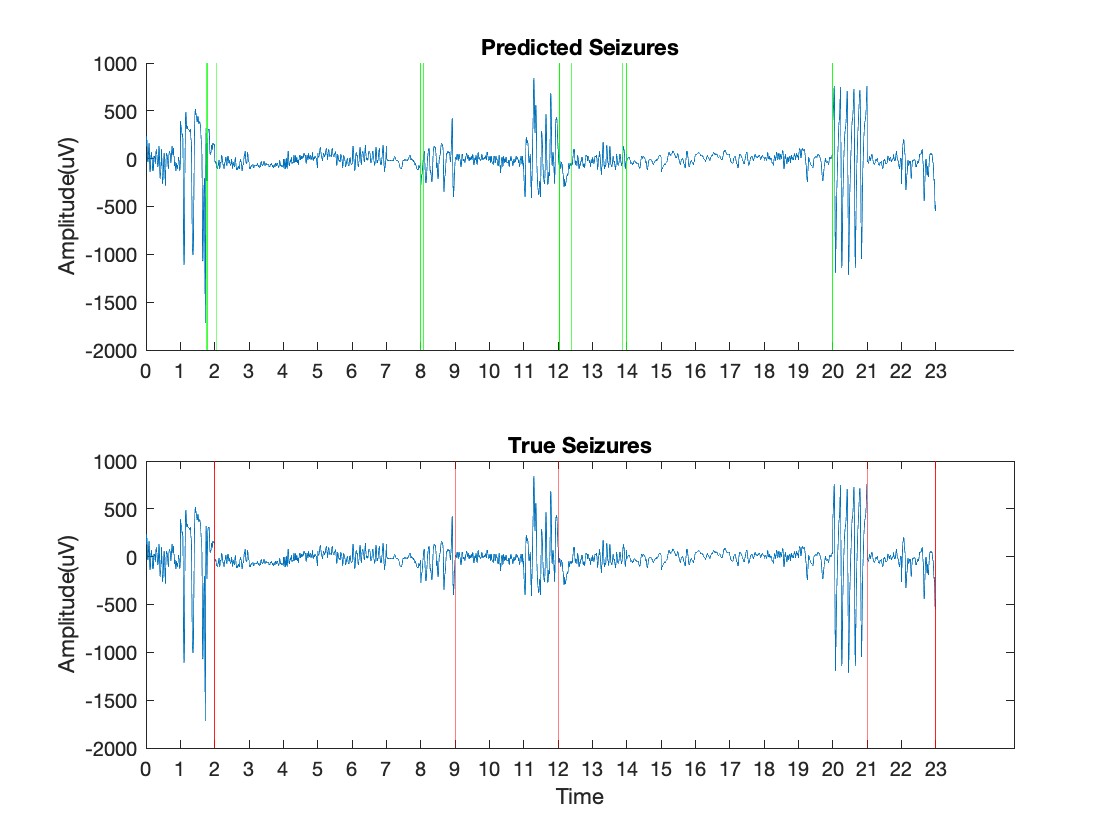} 
    \caption{Seizure detection} 
    \label{fig:detect}
\end{figure}

\begin{figure}[H]
    \centering \includegraphics[width=0.75\columnwidth]{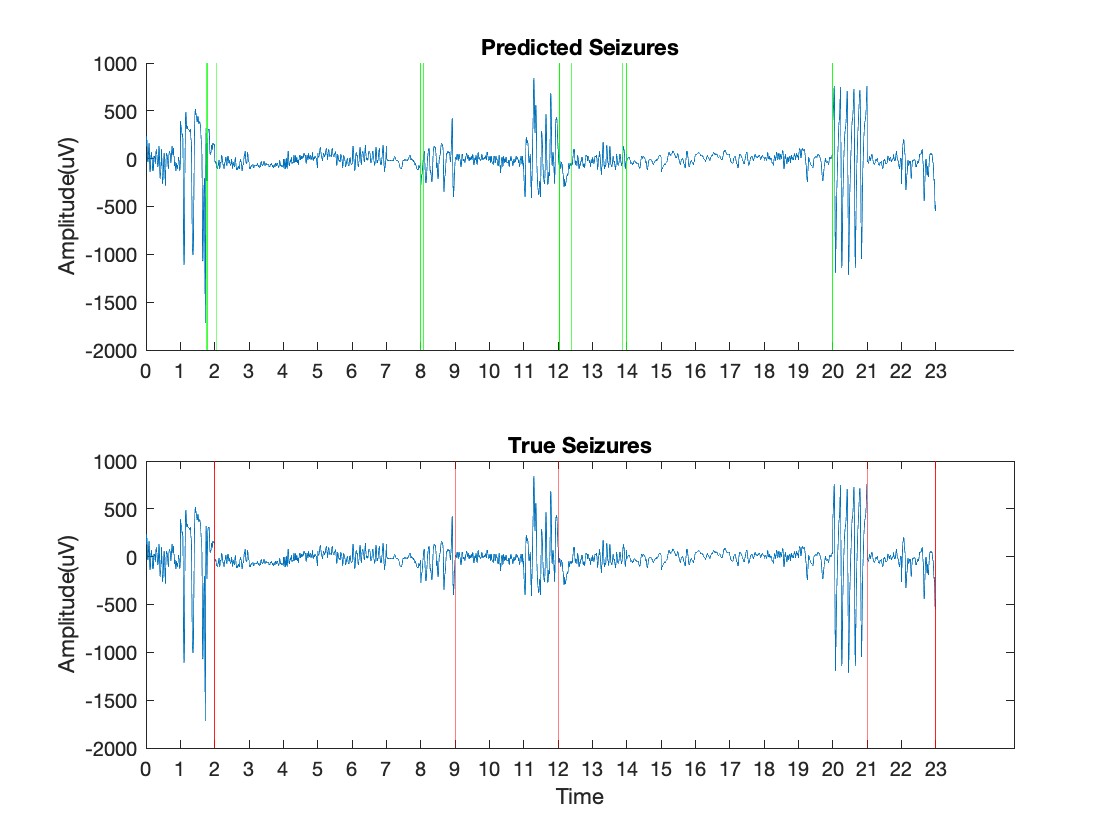} 
    \caption{Seizure detection} 
    \label{fig:detect2}
\end{figure}

\begin{figure}[H]
    \centering \includegraphics[width=0.75\columnwidth]{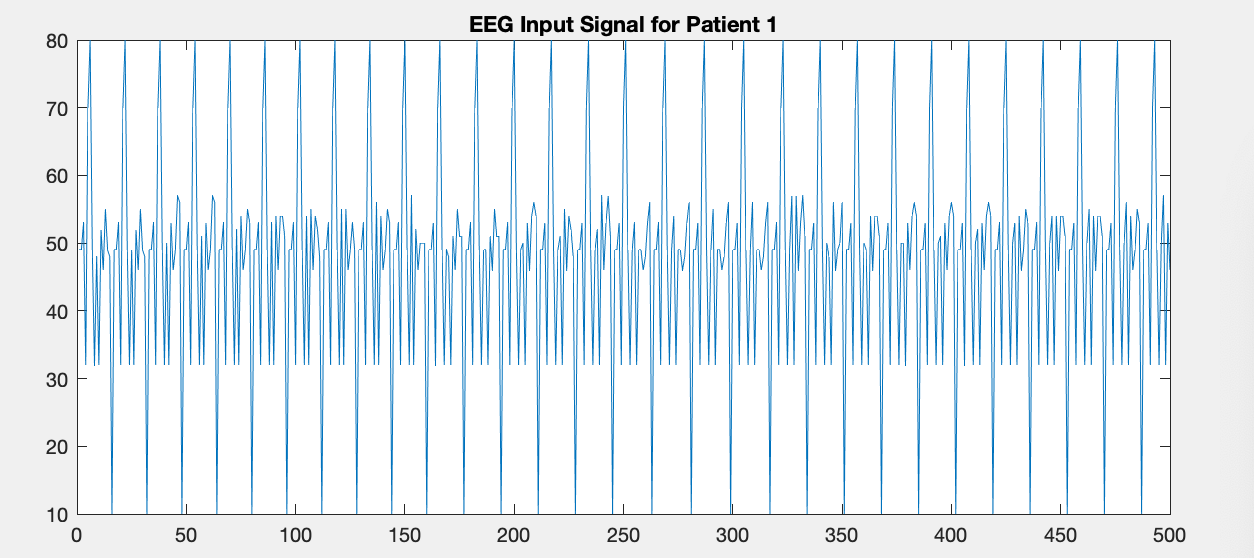} 
    \caption{Seizure detection} 
    \label{fig:eeg_12}
\end{figure}

\begin{figure}[H]
    \centering \includegraphics[width=0.75\columnwidth]{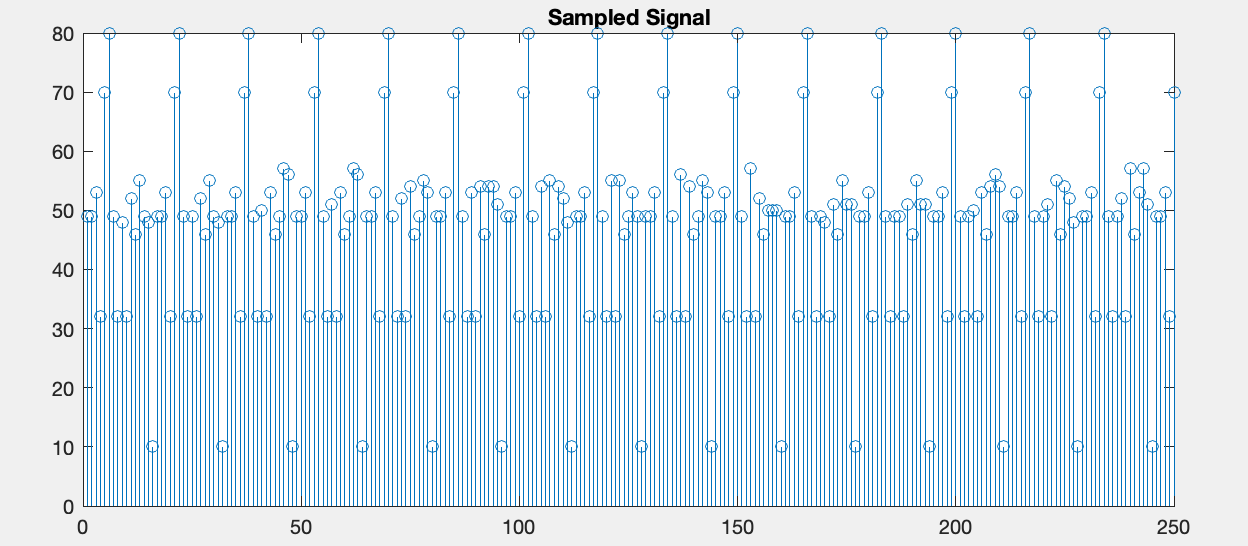} 
    \caption{Seizure detection} 
    \label{fig:eeg_11}
\end{figure}

\begin{figure}[H]
    \centering \includegraphics[width=0.9\columnwidth]{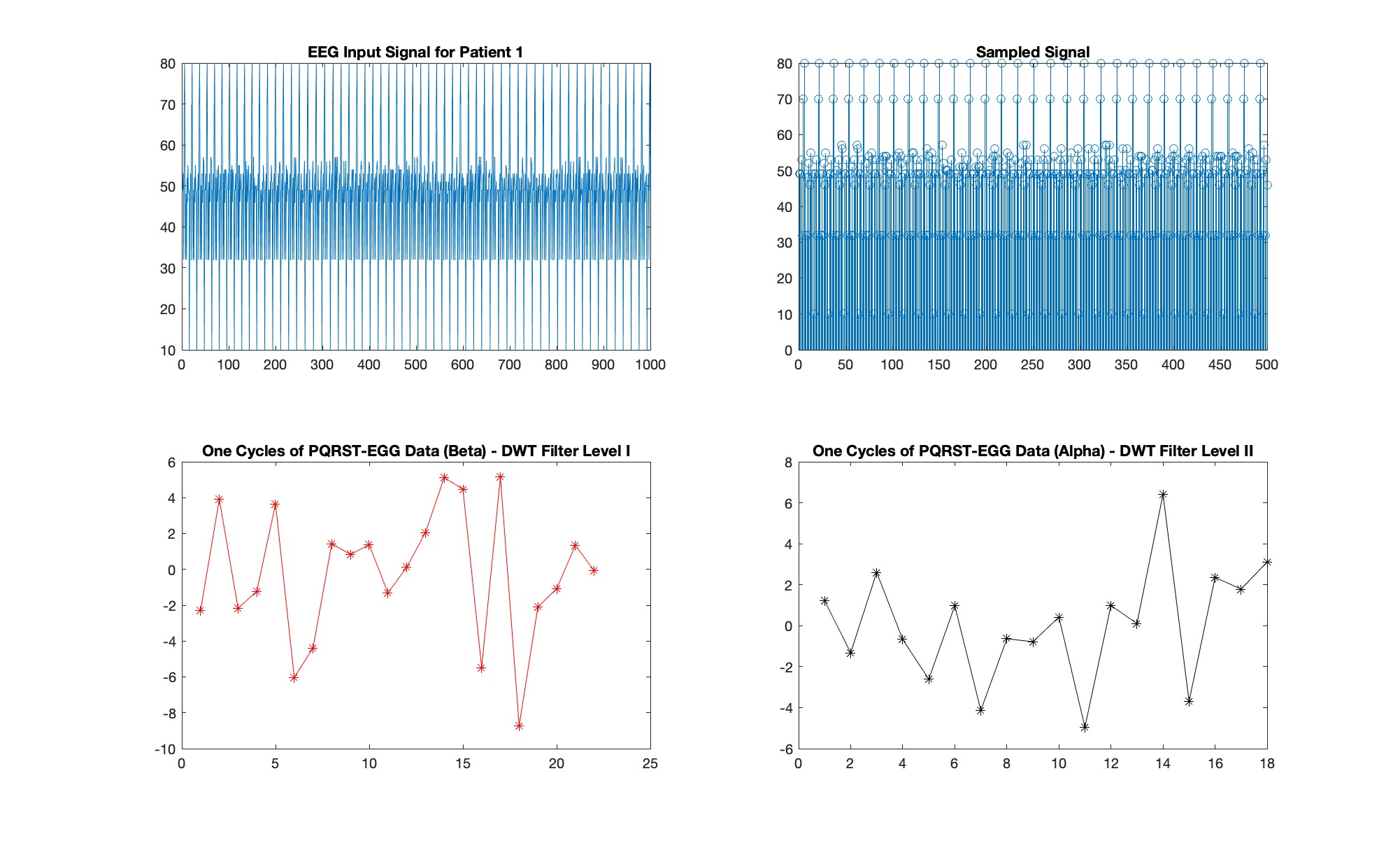} 
    \caption{Extracting alpha and beta channels using db8 wavelet} 
    \label{fig:eeg_1}
\end{figure}

\begin{figure}[H]
    \centering \includegraphics[width=0.9\columnwidth]{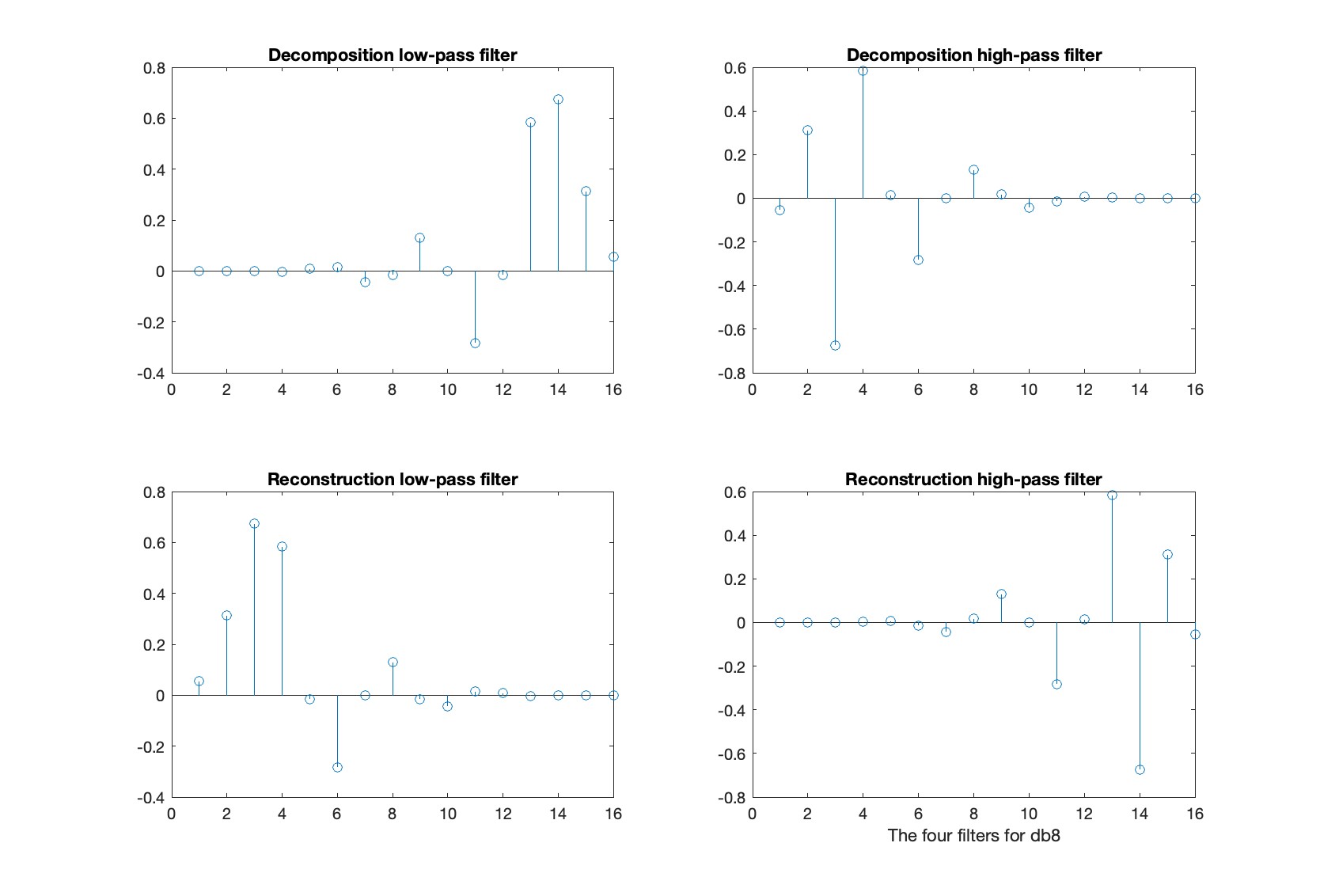} 
    \caption{Extracting alpha and beta channels using db8 wavelet} 
    \label{fig:db8filter}
\end{figure}

Alpha waves can be extracted using the detail 7 coefficient (D7) of the wavelet and the beta wave can be extracted using the detail 6 coefficient (D6).  For instance, in Matlab:
\begin{lstlisting}
    waveletFunction = 'db8';
    [C,L] = wavedec(s(i),8,waveletFunction);
    cD6 = detcoef(C,L,6); 
    cD7 = detcoef(C,L,7); 
    beta = cD6;
    alpha = cD7; 
\end{lstlisting}
    Figure \ref{fig:waveletdec} shows a wavelet decomposition of the EEG signal for 5 levels.
    One can estimate the variance change points of a signal use the wavelet coefficients using the Matlab function $\code{wvarchg}(Y,K)$ where $Y$ is the input signal and $K$ is the number of change points.

\ \ \ Based on the variance changepoints from the detail coefficients from the wavelet decomposition, we can detect seizure point by comparing them along size the true seizure points obtained from the dataset as illustrated in Figure \ref{fig:detect}
    As Figure \ref{fig:detect2} shows, the variance changepoints correctly predicts seizures and matches true seizures at time step 2 and 12, but misses some of the true seizures at timesteps 9, 21, and 23.      
    In Figure \ref{fig:eeg_12} shows a signal of patient from the Epileptic Seizure Recognition Dataset from UC Irvine Machine Learning Repository.\footnote{\url{https://archive.ics.uci.edu/dataset/388/epileptic+seizure+recognition}}
    Figure \ref{fig:eeg_11} shows a downsampling of 2 of the dataset.
    We can extract the alpha and beta channels from this EEG signal, as shown in Figure \ref{fig:eeg_1} by using the 7 detail coefficient (cD7) and 6 detail coefficient (cD6), respectively, from a Daubechies 8 wavelet (db8).  
    We can then use decomposition low-pass and high-pass filters of this signal and reconstruction low-pass and high-pass filters as shown in Figure \ref{fig:db8filter} by using the Matlab \code{wfilters} function.  These four filters are associated with the orthogonal or biorthogonal wavelet \code{wname} which in this case is db8.

\section{Experiment 2: EEG Seizure Recognition Classification}

    \ \ \ Deep neural networks are highly conducive to processing EEG signals since such networks replicate the network of the brain.  Neurons communicate through a combination of chemical neurotransmitters and electrical gradients.  EEG, detects those electrical gradients to provide insight into the activity of the brain.  A deep neural network uses nodes as representations of neurons in the  brains that are connected to each other through weights.  The magnitude of the weights are analogous to synapses

    We use an open-source EEG database from the University of Bonn 
    which is considered a data benchmark for seizure detection.\footnote{\url{https://www.ukbonn.de/epileptologie/arbeitsgruppen/ag-lehnertz-neurophysik/downloads/}}  The dataset is divided into five subsets of ictal scalp EEG signals {F, N, O, Z, and S}, which were collected from 25 subjects.  {O} and {Z} are provided by five healthy (normal) volunteers, with eyes open and closed, respectively, and were collected from scalp surface EEG. Subsets {F, N} and {S} contain EEGs from epileptic patients of which subsets {F, N} and {S} contain EEGs from epileptic patients, of which subsets {F,N} were, respectively, recorded in seizure-free intervals from five patients and subset {S} includes seizure activity recorded from all intracranial sites \cite{Wang:2019}. 

    Epilepsy is a chronic disorder of the brain that affects people of all ages caused by excessive neurological stimuli.  Approximately 50 million people currently live with epilepsy worldwide with close to 80\% living in developing countries. Experienced professionals are able to detect epilepsy by analyzing patterns of the electroencephalogram (EEG). However, in the absence of experts, an automated system is desirable.  
    
    Deep learning networks can automate this process and wavelets can help preprocess and denoise the signals before training. All EEG signals were extracted by the 128-channel amplifier system with an average common reference. Each subset contained 100 samples of EEG signals from 5 subjects. The raw EEG signals were recorded using an international standard 10–20 system with a 173.61 Hz sampling frequency using 12-bit resolution. The age of the subjects ranged from 19 to 60 years. They were all right-handed, and the locations of the epileptogenic foci for each subject were identified by experienced neurologists or epileptologists \cite{Wang:2019}.

\begin{figure}[H]
    \centering \includegraphics[width=0.9\columnwidth]{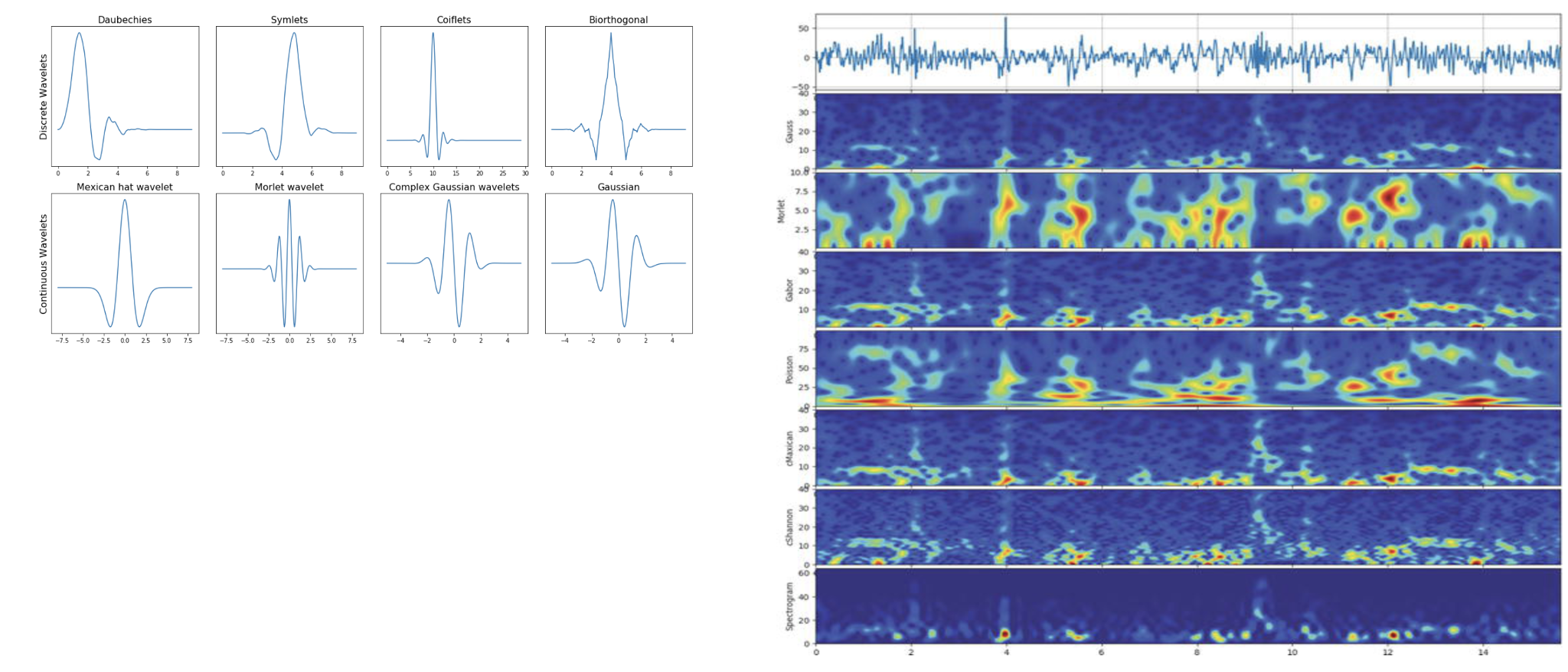} 
    \caption{Illustration of scalograms using different wavelets.  \cite{Taspinar:2018}} 
    \label{fig:scalogram3}
\end{figure}

\begin{figure}
    \centering \includegraphics[width=0.8\columnwidth]{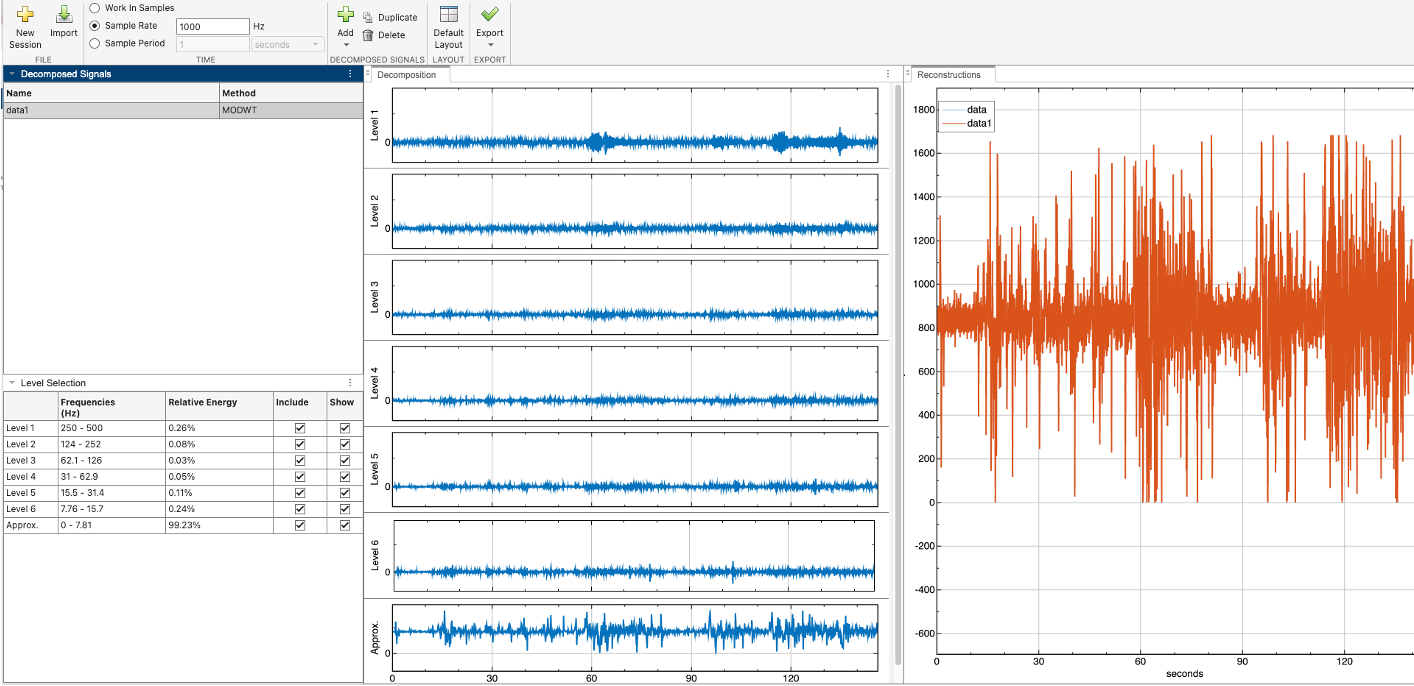} 
    \caption{Matlab Signal Analyzer of EGG Signal} 
    \label{fig:matlab}
\end{figure}

\begin{figure}
    \centering \includegraphics[width=0.8\columnwidth]{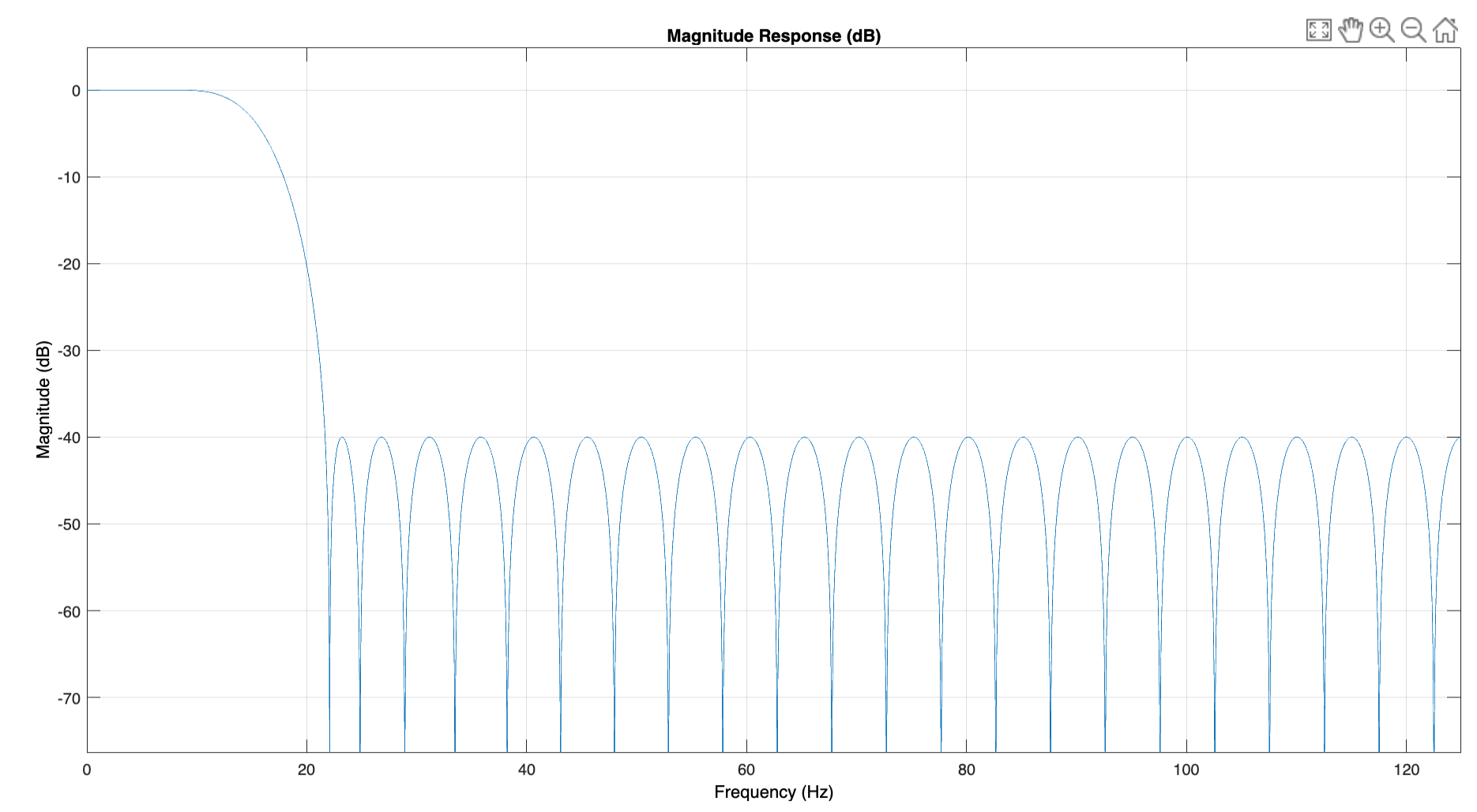} 
    \caption{Magnitude frequency response from wavelet filtering} 
    \label{fig:matlab1}
\end{figure}

    In Python, we use the PyEEG library \cite{Liu:2011} \code{pywt}, PyWavelet, library to analyze the wavelets used to filter and extract features from the EEG signals for non-seizure and seizure subjects. Various features can be extracted from EEG signals including singular value decomposition (SVD) entropy, fisher information, Hurst exponent, relative intensity ratio (RIR), and power spectrum density/intensity \cite{Liu:2011}.  We can convert the signals into scalogram images using the CWT and then input them into a convolutional neural network to extract features.  Figure \ref{fig:scalogram3} shows the scalogram of a signal using eight different types of wavelets: Daubechies, Symlets, Coiflets, Biorthogonal, Mexican hat, Morlet, Complex Gaussian, and Gaussian.
    We use the Matlab Signal Analyzer to visualize the various wavelet decomposition levels of an EGG signal of a patient in the dataset and the magnitude response plot from as illustrated in Figure \ref{fig:matlab} and Figure \ref{fig:matlab1}.
    As shown, the magnitude response creates equidistant notches that have the same amplitude at -40 db.

\begin{figure}[H]
    \centering \includegraphics[width=0.9\columnwidth]{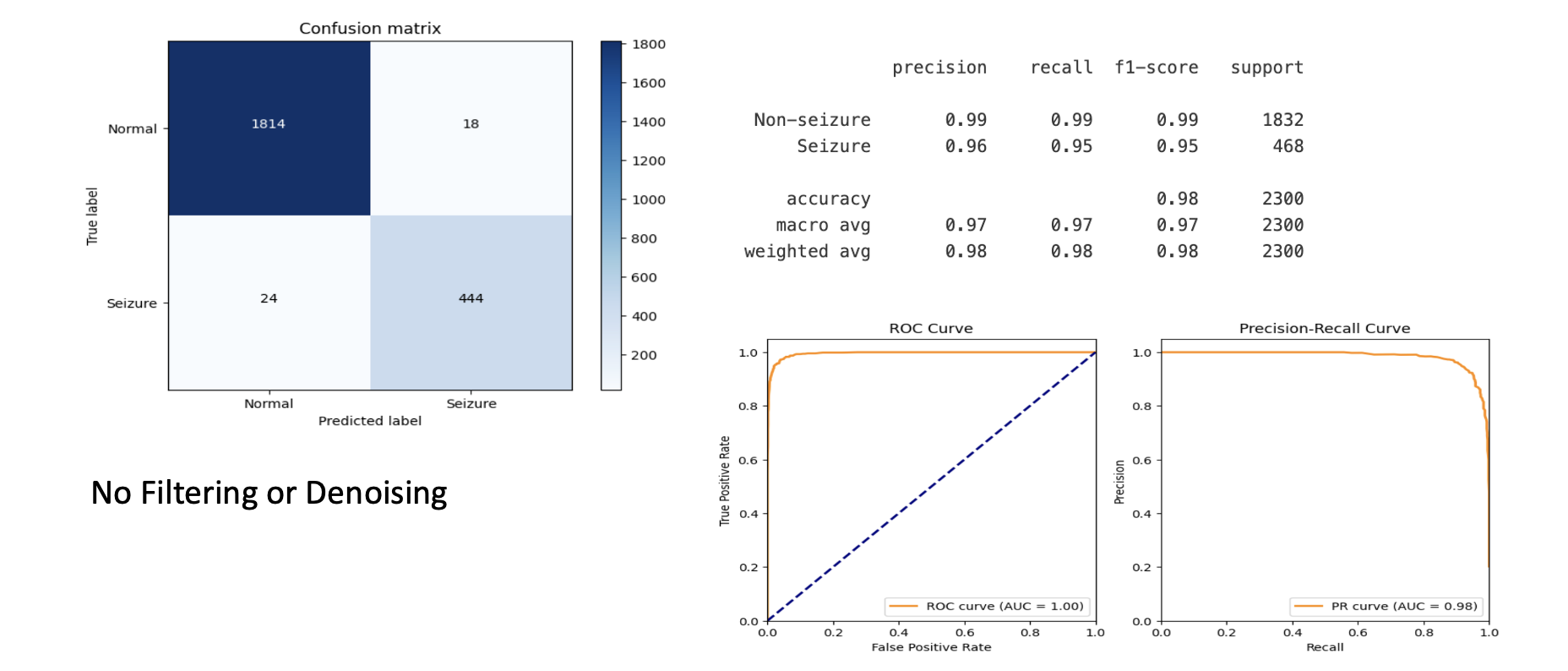} 
    \caption{EEG Classification Results without DFT Denoising} 
    \label{fig:res}
\end{figure}

\begin{figure}[H]
    \centering \includegraphics[width=0.9\columnwidth]{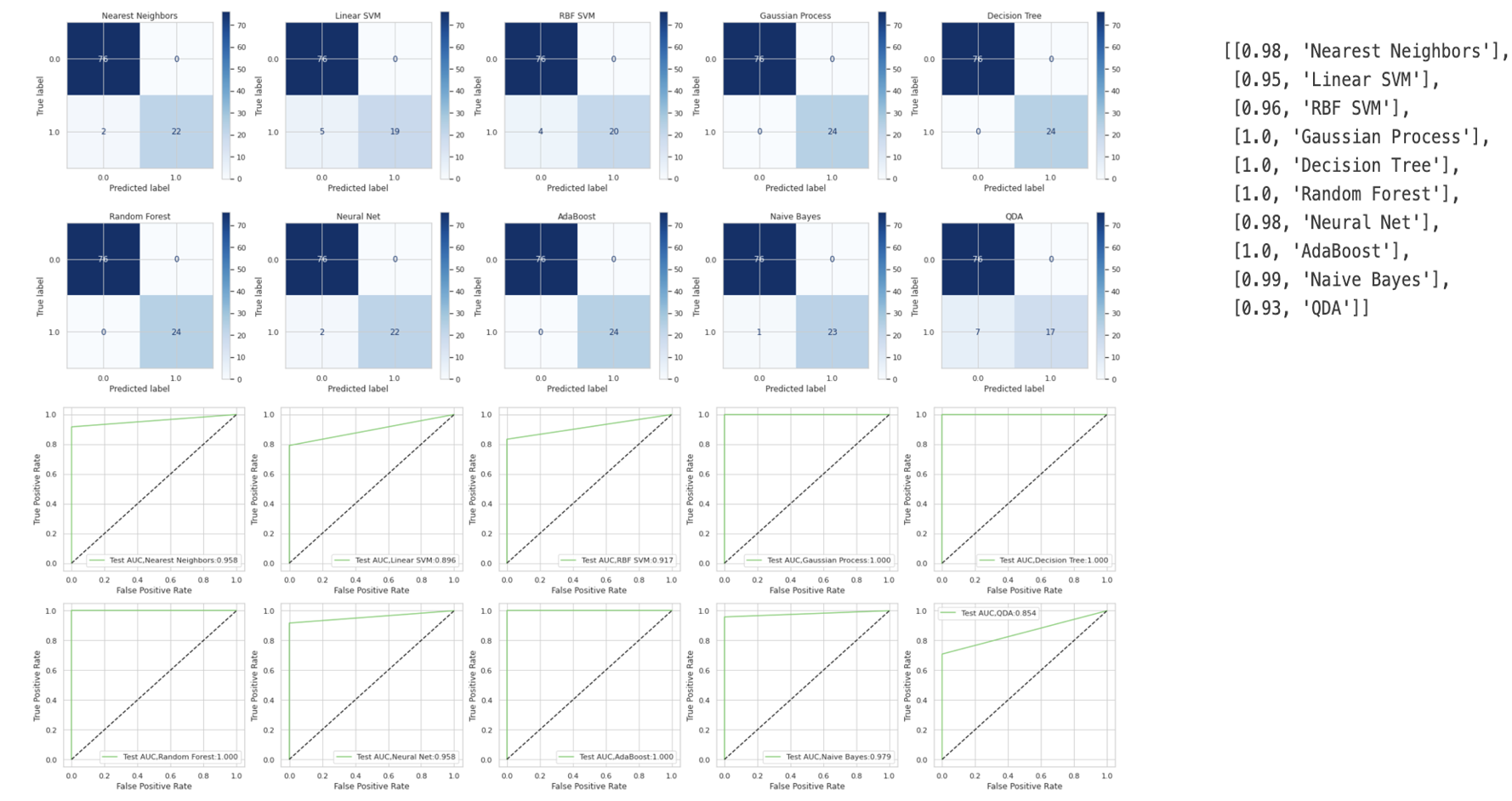} 
    \caption{ROC and Confusion Matrices for Seizure Classification of Various ML Models} 
    \label{fig:res3}
\end{figure}

    Figure \ref{fig:res} shows the the confusion matrix and F1, precision, and recall scores for classification using the DFT using N=1024 points while Figure \ref{fig:res} shows the results with denoising.  As shown, even without denoising, the CNN is able to accurately classify a seizure from a non-seizure signal almost 100\% of the time.   
\begin{figure}[H]
    \centering \includegraphics[width=0.9\columnwidth]{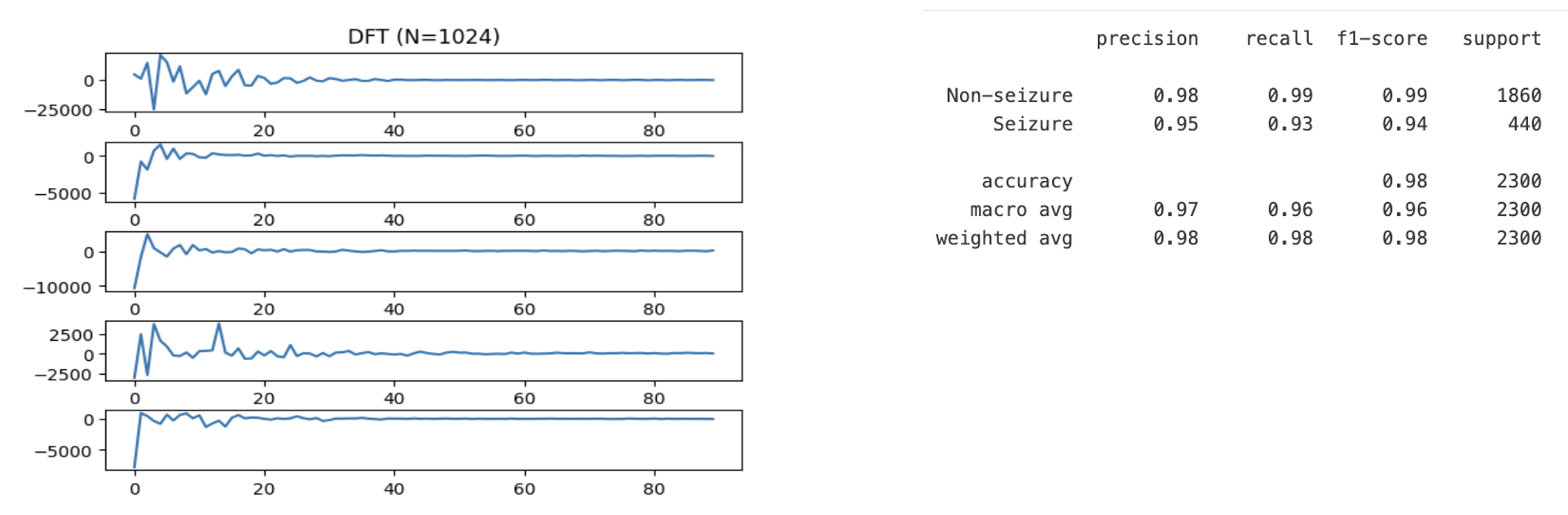} 
    \caption{EEG Classification Results with DFT} 
    \label{fig:res2}
\end{figure} 
    Figure \ref{fig:res3} shows the confusion matrix plots and ROC curves for classification without filtering/denoising using other supervised and unsupervised models including nearest neighbors, linear support vector machines (SVM), Gaussian Processes (GP), decisions trees (DT), random forests (RF),  ANNs, AdaBoost, Naive Bayes, and QDA, respectively.  They also are able to classify seizure from non-seizure signals perfect to near-perfect suggesting that these models can easily handle signal noise in this binary classification.

\section{Experiment 3: Human Activity Recognition (HAR) Signals}

\ \ \ Using the University of Irvine-California (UIC) dataset\footnote{\url{http://archive.ics.uci.edu/dataset/240/human+activity+recognition+using+smartphones}}, we analyze the impact of wavelets for denoising for human activity recognition (HAR).  The experiments to obtain the data were carried out with a group of 30 volunteers within an age bracket of 19-48 years. Each person performed six activities (WALKING, WALKING UPSTAIRS, WALKING DOWNSTAIRS, SITTING, STANDING, LAYING) wearing a smartphone (Samsung Galaxy S II) on the waist. 

Using its embedded accelerometer and gyroscope, data was captured 3-axial linear acceleration and 3-axial angular velocity at a constant rate of 50Hz. The experiments have been video-recorded to label the data manually. The obtained dataset has been randomly partitioned into two sets, where 70\% of the volunteers was selected for generating the training data and 30\% the test data. 

The sensor signals (accelerometer and gyroscope) were pre-processed by applying noise filters and then sampled in fixed-width sliding windows of 2.56 sec and 50\% overlap (128 readings/window). The sensor acceleration signal, which has gravitational and body motion components, was separated using a Butterworth low-pass filter into body acceleration and gravity. The gravitational force is assumed to have only low frequency components, therefore a filter with 0.3 Hz cutoff frequency was used. From each window, a vector of features was obtained by calculating variables from the time and frequency domain. 

\begin{figure}[H]
    \centering \includegraphics[width=0.9\columnwidth]{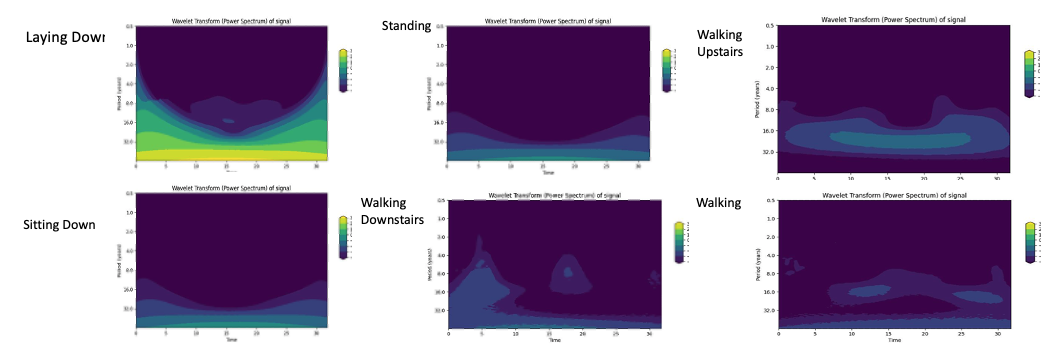} 
    \caption{Scalograms from Wavelet Transform of HAR Signals} 
    \label{fig:harwavelet}
\end{figure}

\begin{figure}[H]
    \centering \includegraphics[width=0.6\columnwidth]{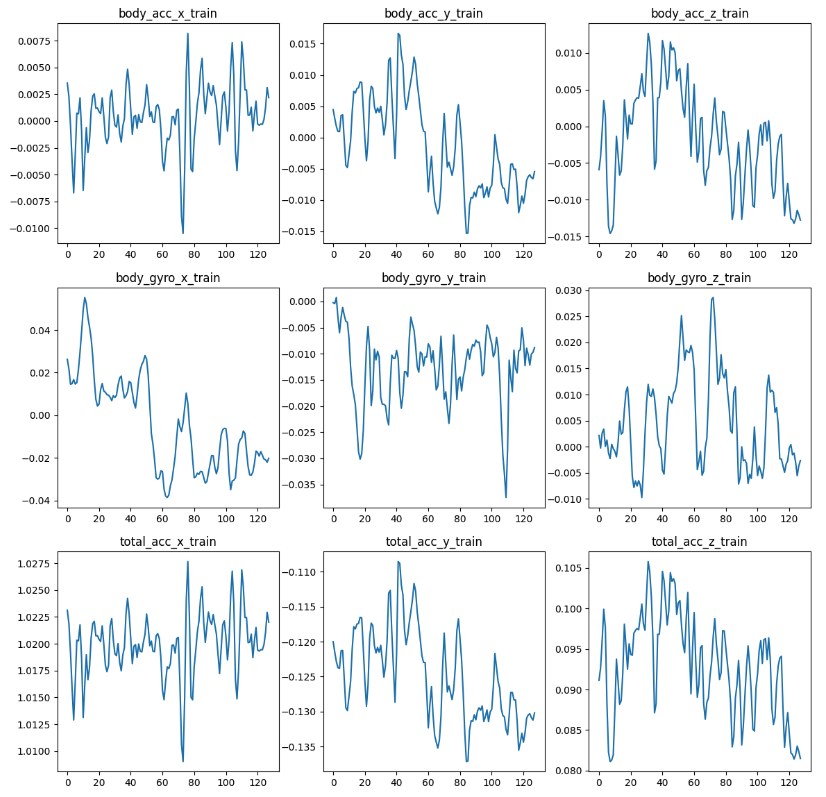} 
    \caption{HAR body and total acceleration and gyroscope sensor measurements} 
    \label{fig:har2}
\end{figure}

\begin{figure}[H]
    \centering \includegraphics[width=0.6\columnwidth]{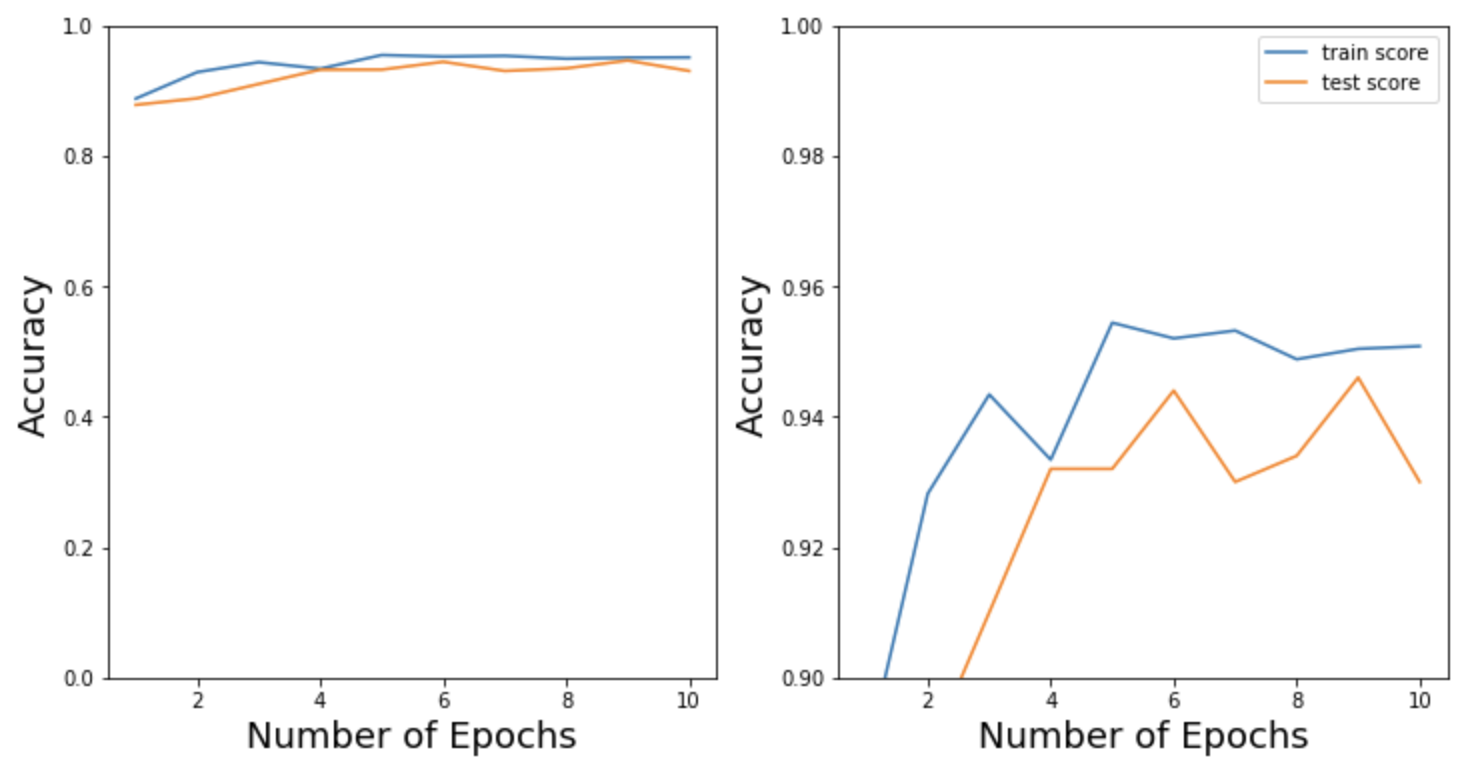} 
    \caption{Training and Test Accuracy. \cite{Taspinar:2020}} 
    \label{fig:wavetrain}
\end{figure}

\begin{figure}[H]
    \centering \includegraphics[width=0.9\columnwidth]{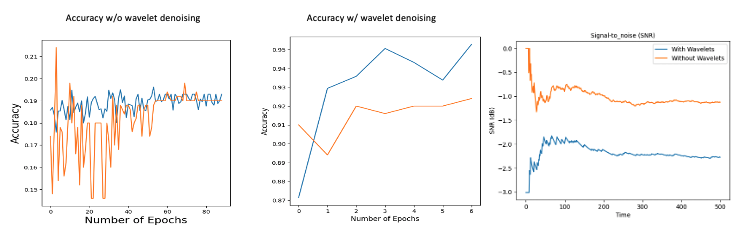} 
    \caption{Comparison of training loss accuracy with and without wavelet transforms} 
    \label{fig:wavelet2}
\end{figure}

Figure \ref{fig:harwavelet} illustrates the power spectrum from wavelet transforms for each of the six activities for a random volunteer.
Figure \ref{fig:har2} illustrates inertial body acceleration, body gyroscope, and total acceleration in the (x,y,z) sensor signals for the second volunteer. 
    A 2D convolutional neural network with two convolutional layers (\code{Conv2D)} using a kernel size of (5,5), a MaxPooling2D layer with a pool size of (2,2) after each Conv2D layer and two dense layers, the first with 1000 units, and the second equal to the number of classes (6) was used for training of the HAR sensor data \cite{Taspinar:2018}.  A batch size of 16 and 10 epochs were used. The image sizes were $127 \times 127$. Training accuracy was 0.959 and the test accuracy was 0.93 using a wavelet transform.  The training loss was 0.111 and the test loss was 0.213.  Figure \ref{fig:wavetrain} shows train and validation (left) and train and test accuracy plots, respectively.  
    Figure \ref{fig:wavelet2} shows a plot of the training accuracy with and without wavelet transforms using 100 epochs using early stopping with a patience of 3.  The training loss converges in only 6 epochs for training with wavelet transformed images but does not stop if wavelet transforms are not used.  
    As shown accuracy substantially improves using accuracy and with far fewer epochs than without wavelets.  Without wavelet transformation, the accuracy is substantially lower (approximately 0.18-0.19).  A plot of the SNR of the signals with and without wavelet transformation shows that the SNR decreases before leveling off while the SNR increases with the wavelet transformation before leveling off.  However, the SNR without wavelet transform is higher for all time steps.  This may be due to the fact that the sensor signal data had been already preprocessed a prior using a low-pass Butterworth filter and thus most of the noise removed.   

\begin{figure}[H]
    \centering \includegraphics[width=0.75\columnwidth]{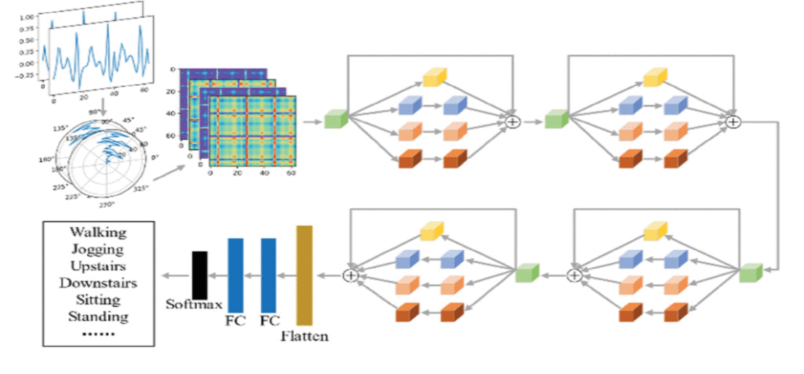} 
    \caption{GAF + CNN Model for HAR Classification. \cite{Xu:2020}} 
    \label{fig:xu}
\end{figure}

\begin{figure}[H]
    \centering \includegraphics[width=0.7\columnwidth]{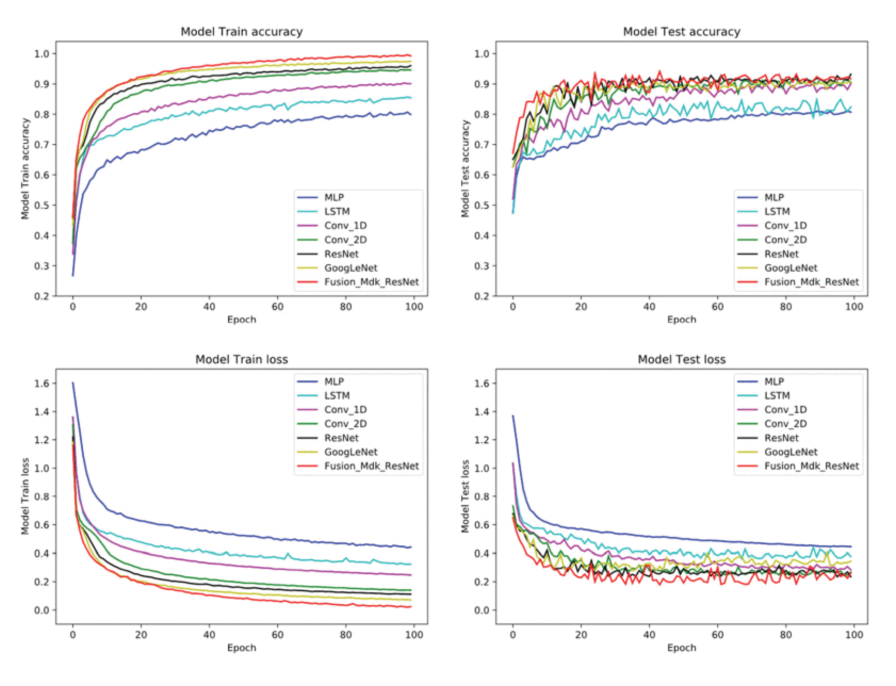} 
    \caption{GAF+CNN Model Results. \cite{Xu:2020}} 
    \label{fig:xu2}
\end{figure}

    Xu, et. al (2020) \cite{Xu:2020} use Gramian Angular Fields (GAF) to first convert 1D HAR signals to 2D images by first converting the time series into polar coordinates and then inputting the images as input into a CNN. A high level overview of their model is shown in Figure \ref{fig:xu}.
    As illustrated in Figure \ref{fig:xu2}, their results show that using GAF yields higher training and test accuracy along with lower training and test loss convergence than CNN models that do not use GAF. 
\section{Experimental Results}

    \ \ \ The experimental results illustrate that wavelet transforms and decomposition of biomedical signals and converting 1D signals to 2D signals via GAF, RP, and/or MTFs algorithms can substantially enhance classification accuracy and loss convergence using deep learning especially when there are more than two classes.   

    In Experiment 1, we analyzed EEG signals.  In the case of EEG seizure recognition, even without denoising or filtering, the classification accuracy of both deep learning and other supervised and unsupervised methods is nearly 100\%. However, in the case of EEG seizure recognition, wavelets do not appear to be useful as with ECGs and HAR signals because it is a binary classification rather than a multi-class problem.  In Experiment 2, we analyzed ECG signals.  We found that wavelet denoising and multi-modal image fusion (MIF) using GAF, RP, and MTF algorithms substantially improves classification accuracy as measured by precision, recall, and F1 scores.  Wavelet transforms and MIF convert 1D signals into 2D images which captures non-stationary dynamics better than 1D time series.  Finally, in Experiment 3, we analyzed HAR signals.  It was shown that using wavelets improves SNR accuracy. 

\section{Conclusion}

\ \ \  Wavelets decmopose signals through a cascade of upsampling (generating approximation coeffients) and downsampling (generating detail coefficients).   A comparison of the classification accuracy between EGGs and ECGs using different datasets was illustrated \cite{Chavez:2020}.  This project sought to (1) to improve the understanding how biomedical signal processing as well as deep learning can be used to quantitatively measure and make sense of biomedical signals for diagnosis and detection of diseases.  We illustrated out wavelet denoising as well as Gramian Angular Fields can improve deep learning classification accuracy by decomposing the signal into an image that contains non-linear dyanmics that cannot be captured in the 2D signal.  We compared the effectiveness of wavelet preprocessing compared to its absence. \

\ \ \ Future work includes analyzing wavelet transforms and comparing their effectiveness with other methods including power spectrum density (PSD) and principal component analysis (PCA)/independent component analysis (ICA).  In addition, analysis of the details of nonlinear features extracted from EEG signals such as approximate entropy (ApEn), Hurst exponent and scaling exponent obtained with detrended fluctuation analysis (DFA) to characterize interictal and ictal states in EEGs and cardiac arrhythmias could help understanding in multiresolution analysis of the signals to detect and identify seizures and heart diseases, respectively.

\newpage
\printbibliography

@article{Wang:2021,
title        = {EEG-GNN: Graph Neural Networks for Classification of
Electroencephalogram (EEG) Signals},
author       = {Demir, A. and Koike-Akino, T. and Wang, Y. and Haruna, M. and Erdogmus, D.},
year         = {2021},
journal      = {arXiv:2106.09135v2 [cs.LG]}
}

@misc{Matlab:2023,
author = {Mathworks},
title = {Time-Frequency Convolutional Network for EEG Data Classification},
url = {https://www.mathworks.com/help/deeplearning/ug/time-frequency-convolutional-network-for-eeg-data-classification.html},
year = {2023},
}

@inproceedings{Anwar:2020,
author = {Anwar, A. and Eldeib, A.},
booktitle = {2020 42nd Annual International Conference of the IEEE Engineering in Medicine \& Biology Society (EMBC)}, 
title = {EEG Signal Classification Using Convolutional Neural Networks on Combined Spatial and Temporal Dimensions for BCI Systems}, 
year = {2020},
pages = {434-437},
}

@article{Craik:2019,
year = {2019},
publisher = {IOP Publishing},
volume = {16},
number = {3},
pages = {031001},
author = {Craik, A. and He, Y. and  Contreras-Vidal, J.},
title = {Deep learning for electroencephalogram (EEG) classification tasks: a review},
journal = {Journal of Neural Engineering},
}

@misc{Edlinger:2023,
title = {EEG Data Processing and Classification with g.BSanalyze Under MATLAB},
author = {Edlinger, G. and Guger, C.},
url = {https://www.mathworks.com/company/newsletters/articles/eeg-data-processing-and-classification-with-gbsanalyze-under-matlab.html},
year = {2023},
}

@misc{Chavez:2020,
author = {Magdalena, C. and Valerdi, A. and María, L. and Zarate, I. and Isaac, D.},
year = {2020}, 
title = {N\&C-TEC: ECG and EEG files}, 
url = {https://data.mendeley.com/datasets/7r4z3p3g4m/1},
doi = {10.17632/7r4z3p3g4m.1}
}

@article{Smigiel:2021,
author = {Smigiel, S., and Palczynski, K. and Ledzinski, D.},
title = {ECG Signal Classification Using Deep Learning Techniques Based on the PTB-XL Dataset}, journal = {Entropy},
volume = {23},
number = {9},
year = {2021}
}

@article{Reyes:2022,
author = {Esqueda-Elizondo, José Jaime and Juárez-Ramírez, Reyes and López-Bonilla, Oscar Roberto and García-Guerrero, Enrique Efrén and Galindo-Aldana, Gilberto Manuel and Jiménez-Beristáin, Laura and Serrano-Trujillo, Alejandra and Tlelo-Cuautle, Esteban and Inzunza-González, Everardo},
title = {Attention Measurement of an Autism Spectrum Disorder User Using EEG Signals: A Case Study},
journal = {Mathematical and Computational Applications},
volume = {27},
year = {2022},
number = {2},
}

@article{Andrezjak:2001,
title = {Indications of nonlinear deterministic and finite-dimensional structures in time series of brain electrical activity: Dependence on recording region and brain state},
author = {Andrzejak, R. and Lehnertz, K. and Mormann, F. and Rieke, C. and David, P. and Elger, C.},
Journal = {Physics Reviews E.},
volume = {64}, 
number = {061907},
year = {2001}
}

@inproceedings{Vaibhav:2023,
author= {Vaibhav, J.
and Tiwari, N.
and Chawla, M.},
editor={Gupta, Deepak
and Khanna, Ashish
and Bhattacharyya, Siddhartha
and Hassanien, Aboul Ella
and Anand, Sameer
and Jaiswal, Ajay},
title= {A Review on EEG Data Classification Methods for Brain-Computer Interface},
booktitle= {International Conference on Innovative Computing and Communications},
year={2023},
publisher= {Springer Nature Singapore},
pages={747-760},
}

@article{Kumar:2022,
title = {Wavelet based machine learning models for classification of human emotions using EEG signal},
journal = {Measurement: Sensors},
volume = {24},
pages = {100554},
year = {2022},
%issn = {2665-9174},
%doi = {https://doi.org/10.1016/j.measen.2022.100554},
url = {https://www.sciencedirect.com/science/article/pii/S266591742200188X},
author = {Shashi {Kumar G S} and Niranjana Sampathila and Tanishq Tanmay},
%keywords = {Electroencephalogram, Discrete wavelet transform, Machine learning, Convolutional neural network, Support vector machine},
%abstract = {Humans have the ability to portray different expressions contrary to the emotional state of mind. Therefore, it is difficult to judge the human's real emotional state simply by judging the physical appearance. Although researchers are working on facial expressions analysis, voice recognition, gesture recognition accuracy levels of such analysis are much less and the results are not reliable. Classifying the human emotions with machine learning models and extracting discrete wavelet features of Electroencephalogram (EEG) is proposed. The EEG data from Database for Emotion Analysis using Physiological signal (DEAP) online datasets is used for analysis and consists of peripheral biological signals as well as EEG recordings. EEG signal is collected from 32 subjects while watching 40 1-min-long music videos. Each video clip is rated by the participants in terms of the level of Valence, Arousal, Dominance. In the proposed work we have considered a significant band of EEG with a reduced frontal electrode (Fp1, F3, F4, Fp2) to get a comparable good result. The accuracy obtained from K- nearest neighbour (KNN), Fine KNN and Support Vector Machine (SVM) are 92.5%, 90% and 90% respectively for Valence, Arousal and Dominance.}
}

@article{Faust:2015,
title = {Wavelet-based EEG processing for computer-aided seizure detection and epilepsy diagnosis},
journal = {Seizure},
volume = {26},
pages = {56-64},
year = {2015},
%issn = {1059-1311},
%doi = {https://doi.org/10.1016/j.seizure.2015.01.012},
url = {https://www.sciencedirect.com/science/article/pii/S1059131115000138},
author = {Faust, O. and Acharya, U. and Adeli, H. and Adeli, A.},
%keywords = {Continuous Wavelet Transform, Discrete Wavelet Transform, Electroencephalogram, Epilepsy},
%abstract = {Electroencephalography (EEG) is an important tool for studying the human brain activity and epileptic processes in particular. EEG signals provide important information about epileptogenic networks that must be analyzed and understood before the initiation of therapeutic procedures. Very small variations in EEG signals depict a definite type of brain abnormality. The challenge is to design and develop signal processing algorithms which extract this subtle information and use it for diagnosis, monitoring and treatment of patients with epilepsy. This paper presents a review of wavelet techniques for computer-aided seizure detection and epilepsy diagnosis with an emphasis on research reported during the past decade. A multiparadigm approach based on the integration of wavelets, nonlinear dynamics and chaos theory, and neural networks advanced by Adeli and associates is the most effective method for automated EEG-based diagnosis of epilepsy.}
}

@article{Arab:2010,
authors = {Arab, M. and Suratgar, A. and AShtiani, A.},
title = {Electroencephalogram signals processing for topographic brain mapping and epilepsies classification},
journal = {Computers in Biology and Medicine},
volume = {40},
number = {9},
pages = {733-739},
year = {2010},
issn = {0010-4825},
}

@article{Xie:2020,
journal = {Xie, Y. and Oniga, S.},
title = {A Review of Processing Methods and Classification Algorithm for EEG Signal},
journal = {Carpathian Journal of Electronic and Computer Engineering},
volume = {13},
number={1},
year = {2020},
pages = {23-29}
}

@article{Xin:2022,
author = {Xin, Q. and Hu, S. and Liu, S. and Zhao, L. and Zhang,Y.},
title = {An Attention-Based Wavelet Convolution Neural Network for Epilepsy EEG Classification},
journal = {IEEE Trans Neural Syst Rehabil Eng},
volume = {30},
pages = {957-966},
year = {2022}
}

@misc{Cui:2016,
author = {Cui, G. and Xia, L. and Liang J. and Tu, M.},
title = {A Classification Algorithm for Epileptic Electroencephalogram Based on Wavelet Multiscale Analysis and Extreme Learning Machine}, 
url = {https://pubmed.ncbi.nlm.nih.gov/29714962/}, 
volume = {33},
number = {6},
pages = {1025-30},
year = {2016}
}

@misc{Brown:2023,
author = {Brown, J.},
title = {Brain waves may predict cognitive impairment in Parkinson's disease},
url = {https://medicine.uiowa.edu/content/brain-waves-may-predict-cognitive-impairment-parkinsons-disease},
year = {2023},
}

@article{Feng:2020,
author = {Li, F. and He, F. and Wang, F. and Zhang, D. and Xia, Y. and Li, X.},
title = {A Novel Simplified Convolutional Neural Network Classification Algorithm of Motor Imagery EEG Signals Based on Deep Learning},
journal = {Applied Sciences},
volume = {10},
year = {2020},
number = {5},
}

@article{Yuan:2011,
title = {Epileptic EEG classification based on extreme learning machine and nonlinear features},
journal = {Epilepsy Research},
volume = {96},
number = {1},
pages = {29-38},
year = {2011},
%issn = {0920-1211},
%doi = {https://doi.org/10.1016/j.eplepsyres.2011.04.013},
url = {https://www.sciencedirect.com/science/article/pii/S0920121111001185},
author = {Yuan, Q. and Zhou, W. and Li, S. and Cai, D.},
}

@article{Kwak:2017,
author = {Kwak, N.S. and Müller, K.R. and Lee, S.W.},
title = {A convolutional neural network for steady state visual evoked potential classification under ambulatory environment},
journal = {PLOS ONE},
year = {2017},
}

@article{Kshirsagar:2018,
author = {Kshirsagar, G.B. and Londhe, N.D.},
title = {Improving performance of Devanagari script input-based P300 speller using deep learning}, 
journal = {IEEE Trans. Biomed. Eng.},
volume = {66}, 
pages = {2992–3005},
year = {2018}
}

@misc{Contreras:2014,
author = {Contreras, M. and Suarez, J.},
title = {EEG Classification with Discrete Wavelet Transforms and Energy Distribution},
institution = {California Polytechnic State University},
url = {https://digitalcommons.calpoly.edu/cgi/viewcontent.cgi?article=1287&context=eesp},
year = {2014}
}

@inproceedings{Avdakovic:1994,
author = {Avdakov, S. and Dizdarevic, K. and Nuhanovic, A.	and Omerhodzic, I.},	
title = {Energy	Distribution of EEG	Signals: EEG Signal	 Wavelet-Neural	Network	Classifier},	
booktitle = {In Proceedings	of	the	Annual	Technical	Conference -Society	of	Plastics	
Engineers},
year = {1994},
pages =	{2865-2867},	
note = {Brookfield,	CT:	Society	of	Plastics Engineers}
}

@book{Subasi:2019,
author = {Subasi, A.},
title = {Practical Guide for Biomedical Signals Analysis Using Machine Learning Techniques: 
A MATLAB Based Approach},
year = {2019},
publisher = {Academic Press : Elsevier}
}

@misc{Jones:2023,
author = {Jones, G.},
title = {Using Machine Learning to Predict Epileptic Seizures from EEG Data},
url = {https://www.mathworks.com/company/newsletters/articles/using-machine-learning-to-predict-epileptic-seizures-from-eeg-data.html},
}

@article{Thir:2021,
author = {Thirugnanam, M. and Pasupuleti, M.},
year = {2021},
month = {06},
title = {Cardiomyopathy -induced arrhythmia classification and pre-fall alert generation using Convolutional Neural Network and Long Short-Term Memory model},
volume = {14},
journal = {Evolutionary Intelligence},
}

@article{Gao:2015,
author = {Gao, W. and Guan, J. and Gao, J. and Zhou, D.},
title = {Multi-ganglion ANN based feature learning with application to P300-BCI
signal classification.},
journal = {Biomed. Signal Process. Control},
year = {2015},
volume = {18},
pages = {127–137}
}

@article{Lu:2016,
author = {Lu, N. and Li, T. and Ren, X. and Miao, H.Y.},
title = {A deep learning scheme for motor imagery classification based on restricted
boltzmann machines},
journal = {IEEE Trans. Neural Syst. Rehabil. Eng.},
year = {2016},
volume = {25},
pages = {566–576}
}

@article{Ma:2017,
author = {Ma, T. and Li, H. and Yang, H. and Lv, X. and Li, P. and Yao, D},
title = {The extraction of motion-onset VEP BCI features based on deep
learning and compressed sensing.},
journal = {J. Neurosci. Methods},
year = {2017},
volume = {77},
number = {275},
pages = {80–92}
}

@article{Müller:2008,
author = {Muller, K.R. and Tangermann, M. and Dornhege, G. and Krauledat, M. and Curio, G. and Blankertz, B.}, 
title = {Machine learning for real-time single-trial EEG-analysis: From brain–computer interfacing to mental state monitoring},
journal = {J. Neurosci. Methods},
year = {2008}, 
volume = {167},
pages = {82–90}
}

@article{Acharya:2005,
author = {Acharya, R. and Faust, O. and Kannathal, N. and Chua, T. and Laxminarayan, S.}, 
title = {Non-linear analysis of EEG signals at various sleep stages},
journal = {Comput. Methods Programs Biomed.}, 
year = {2005},
volume = {80},
pages = {37–45},
}

@article{Tang:2020,
author = {Tang, X. and Li, W. and Li, X. and Ma, Z. and Dang, X.},
title = {Motor imagery EEG recognition based on conditional optimization empirical mode decomposition and multi-scale convolutional neural network},
journal = {Expert Syst. Appl.},
year = {2020},
volume = {149},
pages = {113285}
}

@article{Lotte:2018,
year = {2018},
month = {Apr},
publisher = {IOP Publishing},
volume = {15},
number = {3},
pages = {031005},
author = {F Lotte and L Bougrain and A Cichocki and M Clerc and M Congedo and A Rakotomamonjy and F Yger},
title = {A review of classification algorithms for EEG-based brain–computer interfaces: a 10 year update},
journal = {Journal of Neural Engineering},
}

@article{Kateb:2023,
author = {Petrossian, G. and Kateb, P. and Miquet-Westphal, F. and Cicoira, F.},
title = {Advances in Electrode Materials for Scalp, Forehead, and Ear EEG: A Mini-Review},
journal = {ACS Applied Bio Materials},
volume = {6},
number = {8},
pages = {3019-3032},
year = {2023}
}

@misc{ecgwaves:2023,
author = {ecgwaves.com},
title = {https://ecgwaves.com/ecg-qrs-complex-q-r-s-wave-duration-interval/},
year = {2023}
}

@misc{rwave:2023,
author = {Matlab},
title = {R Wave Detection in the ECG},
year = {2023},
url = {https://www.mathworks.com/help/wavelet/ug/r-wave-detection-in-the-ecg.html}
}

@article{Almousa:2023,
author = {Al-mousa, A. and Banissa, J. and Hashem, T. and Ibraheem, T.},
title = {Enhanced electrocardiogram machine learning-based classification with emphasis on fusion and unknown heartbeat classes},
year = {2023},
journal = {Digital Health},
url = {https://www.ncbi.nlm.nih.gov/pmc/articles/PMC10353025/}
}

@ARTICLE{Zaid:2020,
  author={Alyasseri, Zaid Abdi Alkareem and Khader, Ahamad Tajudin and Al-Betar, Mohammed Azmi and Abasi, Ammar Kamal and Makhadmeh, Sharif Naser},
  journal={IEEE Access}, 
  title={EEG Signals Denoising Using Optimal Wavelet Transform Hybridized With Efficient Metaheuristic Methods}, 
  year={2020},
  volume={8},
  number={},
  pages={10584-10605},
  doi={10.1109/ACCESS.2019.2962658}
}

@article{Alyasseri:2018,
author = {Alyasseri, Z. and Kader, A. and Al-Betar, J. and Pap, P. and Alomari, A.},
title = {EEG feature extraction for person identification using wavelet decomposition and multi-objective flower pollination algorithm},
journal= {IEEE Access},
volume = {6},
year = {2018},
pages = {76007-760024},
}

@article{Gopan:2022,
author = {Shah, D. and Gopan, G. and Sinha, N. },
title = {An investigation of the multi-dimensional (1D vs. 2D vs. 3D) analyses of EEG signals using traditional methods and deep learning-based methods},
journal = {Frontiers in Signal Processing},
volume = {2},
year = {2022}
}

@article{Rizwan:2022,
author = {Rizwan, A and Priyanga, P. and Abualsauod, E. and Zafrullah, S. and Serbaya, S. and Halifa, A.},
title = {A Machine Learning Approach for the Detection of QRS Complexes in Electrocardiogram (ECG) Using Discrete Wavelet Transform (DWT) Algorithm},
journal = {Computational Inelligence in Neuroscience},
year = {2022},
}

@article{Pan:1985,
author = {Pan, J. and Tompkins, W.}, 
title = {A Real-Time QRS Detection Algorithm. IEEE Transactions on Biomedical Engineering},
journal = {IEEE Transactions on Biomedical Engineering},
volume = {32},
number = {3},
pages = {230-236},
}

@misc{Sharma:2021,
author = {Sharma, K.},
publisher = {GitHub},
journal = {GitHub repository},
url  = {https://github.com/antimattercorrade/Pan_Tompkins_QRS_Detection},
year = {2021},
}

@article{Wang:2019,
author = {Wang, X. and Gong G and Li, N.},
title = {Automated Recognition of Epileptic EEG States Using a Combination of Symlet Wavelet Processing, Gradient Boosting Machine, and Grid Search Optimizer},
journal = {Sensors},
volume = {19},
number = {2},
year = {2019},
doi = {10.3390/s19020219},
url = {https://www.ncbi.nlm.nih.gov/pmc/articles/PMC6359608/}
}

@article{Liu:2011,
author = {Liu, X.  and Zhang, C.},
title = {PyEEG: An Open Source Python Module for EEG/MEG Feature Extraction},
journal = {Computational Intelligence and Neuroscience},
doi = {https://doi.org/10.1155/2011/406391},
year = {2011}
}

@misc{Taspinar:2018,
author = {Taspinar, A.},
title = {A guide for using the Wavelet Transform in Machine Learning},
url = {https://ataspinar.com/2018/12/21/a-guide-for-using-the-wavelet-transform-in-machine-learning/},
year = {2018}
}

@article{Ahmad:2021,
author = {Ahmad, Z. and Tabassum, A. and Gaun, L. and Khan, N.},
title = {ECG Heartbeat Classification Using Multimodal Fusion},
journal = {IEEE Access},
doi = {10.1109/ACESS.2021.3097614},
volume = {9},
pages = {100615-100626},
year = {2021},
}

@article{Xu:2018,
author = {Xu, S.S. and Mak, M.W. and Cheung, C.C.},
title = {Towards end-to-end ECG classification with raw signal extraction and deep neural networks},
journal = {IEEE J. Biomed. Health Information},
year = {2018}, 
volume= {23}, 
pages = {1574–1584},
}

@misc{Taspinar:2020,
author = {Taspinar, A.},
title= {Classification of ECG signals using the Discrete Wavelet Transform and Gradient Boosting},
url = {https://github.com/taspinar/siml},
year = {2020},
}

@misc{Sirjani:2022,
author = {Sirjani, M.},
title = {EEG Based Epilepsy Detection},
url = {https://github.com/msadeqsirjani/eeg-based-epilepsy-detection},
year = {2022}
}

@article{Xu:2020,
author = {Xu, H. and Li, J. and Yuan, H. and Liu, Q. and Fan, S. and Li, T. and Sun, X.},
title = {Human Activity Recognition Based on Gramian Angular Field and Deep Convolutional Neural Network},
journal = {IEEE Access},
volume = {8}, 
year = {2020},
doi = {DOI: 10.1109/ACCESS.2020.3032699},
pages = {199393-199405}
}

@inproceedings{Wang:2015,
author = {Wang, Z. and Oates, T.},
title = {Imaging time-series to improve classification and imputation}, 
booktitle = {Proc. 24th Int. Joint Conf. Artif. Intell. (IJCAI}, 
pages = {3939-3945}, 
year = {2015},
}

@article{Wang2:2015,
author = {Wang, Z. and Oates, T.},
title = {Spatially Encoding Temporal Correlations to Classify Temporal Data Using Convolutional Neural Networks},
year = {2015},
journal = {arXiv:1509.07481},
}

@misc{Tabassum:2020,
author = {Tabassum, A.},
title = {Heartbeat Classification},
url = {https://github.com/atabas/Heartbeat-Classification},
year = {2020},
}

@article{Zhang:2022,
author = {Zhang, H. and Liu, C. and Tang, F. and Li, M. and Zhang, D. and Xia, L. and Zhao, N. and Li, S. and Crozier, S and Xu, W. and Liu, F.},
title = {Cardiac Arrhythmia classification based on 3D recurrence plot analysis and deep learning},
journal = {Frontiers in Physiology},
year = {2022},
doi = {https://doi.org/10.3389/fphys.2022.956320}
}

@misc{Dominguez:2019,
author = {Dominguez, M.},
title= {Electrocardiogram (ECG)},
url  = {https://step1.medbullets.com/cardiovascular/108017/electrocardiogram-ecg},
year = {2019}
}

@misc{Lim:2018,
author = {Lim, R.},
title = {QRS Detection and ECG Classifier},
year = {2018}
}

@article{masud:2018,
author = {Masud, A.},
journal = {International Journal of Industrial Electronics and Electrical Engineering (IJIEEE)},
number = {7 (Part 2)},
pages = {39-43},
title = {Real Time P, QRS AND T Wave Detection By QRS Matched Filter Method},
url = {http://www.ijieee.org.in/volume.php?volume_id=482},
volume = {6},
year = {2018},
}

@misc{Abbas:2017,
author = {Abbas, M.},
title = {Disease detection through QRS complex interval in ECG signals},
url = {https://github.com/mohammadzainabbas/ECG-QRS/tree/master},
year = {2017}
}

\end{document}